\documentclass[iop]{emulateapj}
\usepackage{url}
\usepackage{hyperref}

\def\references{\bibliographystyle{/Users/schryver/tex/apj}\bibliography{/Users/schryver/book/references/ref_karel}}

\def\ga{\mathrel{\mathchoice {\vcenter{\offinterlineskip\halign{\hfil
$\displaystyle##$\hfil\cr>\cr\sim\cr}}}
{\vcenter{\offinterlineskip\halign{\hfil$\textstyle##$\hfil\cr>\cr\sim\cr}}}
{\vcenter{\offinterlineskip\halign{\hfil$\scriptstyle##$\hfil\cr>\cr\sim\cr}}}
{\vcenter{\offinterlineskip\halign{\hfil$\scriptscriptstyle##$\hfil\cr>\cr\sim\cr}}}}}

\shorttitle{Large-scale magnetic couplings between solar coronal events}
\shortauthors{Schrijver et al.}

\begin{document}
%\warningoverprint{Submitted to ApJL}
%Accepted for ApJL \ldots}

\title{Pathways of large-scale magnetic couplings between solar coronal events}

\author{Carolus J. Schrijver$^1$, Alan M. Title$^1$, Anthony R. Yeates$^2$, and Marc L. DeRosa$^1$}
\affil{$^1$ Lockheed Martin Advanced Technology Center, 3251 Hanover Street, Palo Alto, CA 94304, USA \\
$^2$ Department of Mathematical Sciences, Durham University, Science Laboratories, South Road, Durham DH1 3LE, UK}

\begin{abstract}
The high-cadence, comprehensive view of the solar corona by SDO/AIA shows many events that are widely separated in space while occurring close together in time. In some cases, sets of coronal events are evidently causally related, while in many other instances indirect evidence can be found. We present case studies to highlight a variety of coupling processes involved in coronal events.  We find that physical linkages between events do occur, but concur with earlier studies that these couplings appear to be crucial to understanding the initiation of major eruptive or explosive phenomena relatively infrequently.  We note that the post-eruption reconfiguration time scale of the large-scale corona, estimated from the EUV afterglow, is on average longer than the mean time between CMEs, so that many CMEs originate from a corona that is still adjusting from a previous event.  We argue that the coronal field is intrinsically global: current systems build up over days to months, the relaxation after eruptions continues over many hours, and evolving connections easily span much of a hemisphere. This needs to be reflected in our modeling of the connections from the solar surface into the heliosphere to properly model the solar wind, its perturbations, and the generation and propagation of solar energetic particles. However, the large-scale field cannot be constructed reliably by currently available observational resources. We assess the potential of high-quality observations from beyond Earth's perspective and advanced global modeling to understand the couplings between coronal events in the context of CMEs and solar energetic particle events. 
\end{abstract}


%\clearpage
\section{Introduction}\label{sec:introduction}
The Atmospheric Imaging Assembly \citep[AIA, ][]{aiainstrument} on the Solar Dynamics Observatory\citep[SDO,][]{2012SoPh..275....3P} provides continuous full-disk observations of the solar chromosphere and corona in seven extreme-ultraviolet (EUV) channels, spanning a temperature range from some 20,000\,K to in excess of 20\,MK \citep{aiacalibration}. The 12-second cadence of the image stream with 4096 by 4096 pixel images at 0.6\,arcsec/pixel provides unprecedented views of the various phenomena that occur within the evolving solar outer atmosphere.

With the full-disk coverage and the multitude of events going on on the Sun at any given time, it is not surprising that many events are seen to occur synchronously or near-synchronously. In many cases, this (near-)synchronicity is a matter of chance.
During the 976-day time interval from the start of the SDO prime mission (2010/05/01) through the last of the searched dates for this study (2012/12/31) the NOAA/GOES logs contained 2881 flares of class C1 or above, or on average 3.0 flares per day. The study by \citet{robbrecht+etal2009} counts typically some 4 coronal mass ejections for an average day. For a causal link via an Alfv{\'e}nic signal, we can take a typical time delay $\Delta\tau$ for a signal to travel a distance $d$ around the solar circumference with a characteristic coronal Alfv{\'e}n speed $v_{\rm A}$. For events separated by, e.g., 90$^\circ$, $d=2\pi R_\odot /4$ and with $v_{\rm A}= 500$\,km/s, we find that $\Delta\tau = 2200$\,s, or about 1/40-th of a day. With 3 flares and 2 CMEs a day on the Earth-facing hemisphere, each of which can last for up to multiple hours, it is not surprise that many events in the SDO/AIA data are seen to overlap in time (quantitative estimates are provided in Section~\ref{sec:discussion}), 
Overlap or proximity in time is not an adequate distinguishing criterion for inferring causal links between events on the Sun. 

In order to assess the importance of causal linkages in the triggering of near-synchronous events, we first must identify and classify the types of pathways that may connect them. To that end, we have reviewed many near-synchronous events in SDO/AIA observations, and here present a selection of those to illustrate the causal linkages. With the available present-day observations, we can see that 
many of the events observed with SDO/AIA either reveal direct magnetic connections between near-synchronously flaring or erupting regions, while others are highly suggestive of it. 

The idea of causal linkage between flaring in different regions goes back to \citet{richardson1936}. \citet{richardson1951}, following up on his initial report, noted that ``the formal statistical results as well as the visual impression conveyed from inspection of the [Ca\,II\,K] photographs suggests that some form of coupling may exist between widely separated spot groups'', but with the data at hand concluded that ``the question must still be regarded as open.''  Since then, multiple studies were published on the possible linkages not only between flares but also between flares and filament activations and even between sets of filament activations; \citet{1966ApJ...145..224D}, who list many of these early studies, differentiate the linkage between relatively distant regions from another form of linkage, namely that of multiple flares occurring sequentially in the same region, which includes the subset of homologous flares \citep[with a much-larger scale equivalent in repeating pseudo-streamer blowouts, e.g.,][]{2012arXiv1212.6677L}.  

Among the possible scenarios for linkage between solar impulsive events that were proposed already early on were fast-moving energetic particles and also shock waves traveling at up to 2000\,km/s. Evidence for the existence and possible role of these was found at radio and microwave wavelengths \citep[e.g.,][]{1969PASAu...1..181W,1970SoPh...13..227F}. Thermal energy input conducted via direct field connections was also proposed as a possible causal agent in sympathetic activity \citep[e.g.,][]{2000SoPh..195..135C}.

Larger data bases, some supported by space-based observations of coronal connections, led to statistical studies in the context of connection patterns in the coronal magnetic field. \citet{1976SoPh...48..275F} and \citet{1990A&A...228..513P}, for example, following similar studies and conclusions by others cited in their work, used those statistical methods on samples of events to conclude that sympathetic flaring was at most a weak phenomenon, although apparently significant for regions in close proximity of each other (which those studies found to be closer than $30^\circ$ and $35^\circ$, respectively). 

Suggestions of causal linkage between events are not limited to the flare-flare or flare-filament events, but also include couplings of filament eruptions. For example, \citet{2011ApJ...738..179J} discuss the possibility of two quiet-Sun filament eruptions occurring in the wake of an active-region eruption, with the coupling agent being the field deformation by an eruption (often with a signature ``coronal dimming'') that is instrumental in causing other field configurations that are either connected to it  or that lie contained within it to lose their stability. Another such pair of related quiet-Sun filament eruptions, along with other flaring activity, was discussed by \citet{schrijver+title2010}. A magnetic configuration of three adjacent flux ropes was subsequently modeled by \citet{2011ApJ...739L..63T}, whose detailed MHD work illustrated how stretched field by one eruption can destabilize one or more other flux-rope configurations nested within a common overlying field. Such linkages, particularly those involving fairly high field, are often not directly observable but can be inferred from models, as was done in the studies mentioned in this paragraph, and in the work by, for example, \citet{2012ApJ...745....9Y} and \citet{2012ApJ...750...12S}.

Despite the many studies cited above and referenced within those, the phenomenon of sympathetic activity in the solar corona remains elusive. Yet, in the context of forecasting solar activity, the influence of adjacent or distant regions on the loss of stability of a given region needs to be understood. This is particularly important for the development and  propagation of CMEs and the resulting particle events and geomagnetic storms: given the frequency of eruptive events on the Sun, many CMEs are composite events, but understanding their makeup from different events with either physical connections or chance coincidences is important both to the  interpretation and the forecast of any heliospheric event under study. 

The complete coverage of the Earth-facing hemisphere of the solar surface and corona by SDO, supported by far-side observations from the STEREO spacecraft \citep{2008SSRv..136....5K}, complemented by the STEREO and SOHO/LASCO \citep{brueckner+etal1995} coronagraphs and full-sphere coronal field modeling \citep[such as work by][that we use in the present study]{schrijver+derosa2002b,yeates+etal2008} enable a comprehensive empirical assessment of the linkages that exist in the solar corona that may play a role in sympathetic activity.
In view of the mixture of positive and negative findings in the literature of the significance of causal linkages between explosive or eruptive events on the Sun, the aim of this study is not to quantify the frequency of sympathetic couplings, but to assess the evidence for, and to discuss examples of, any of the proposed causal pathways by which couplings may occur. 

In selecting the events discussed here, we reviewed much of the SDO/AIA data for the period from 2010/05/01 through 2012/12/31. Based on the review of those observations, we discuss evidence for four fundamentally distinct plausible causal pathways:  (evolving) direct magnetic connections, waves or propagating disturbances, distortions of and reconnection with overlying field by the eruption of one or more flux ropes elsewhere in the corona, and evolving indirect connections.  One aim of this study is to present evidence for each of these pathways in at least one well-observed case. Our other aim is to illustrate why it is proving difficult to assess the prevalence of sympathetic couplings in the causes of space weather and to point towards future opportunities to alleviate the difficulties in the study of this now nearly 80-year old problem. 

Section~\ref{sec:data} describes the various data sets used in this study and the selection criteria for initial review of candidate data sets. Section~\ref{sec:casestudies} reviews the most illuminating cases that support causal linkage of solar events or clearly illustrate a particular difficulty in establishing whether events are synchronous by chance or by interaction. In reviewing each of the events, we typically look at events over a full 24-h period and sometimes up to several days. After reviewing Section~\ref{sec:data} the reader may choose to jump to the final two sections, where we discuss our findings in a summarizing  Section~\ref{sec:conclusions} and in the concluding Section~\ref {sec:discussion}, before reviewing the detailed case studies in Section~\ref{sec:casestudies}. To aid in the
review of the supporting images and movies, we created an on-line Table \url{http://www.lmsal.com/forecast/STYD.html} and include direct links throughout the manuscript shown by a superscript ``S'' followed by a number.

\section{Data}\label{sec:data}
The primary data source for this study is the complete archive of SDO/AIA observations. We selected candidate events using several different approaches. First, we created 30-minute summed images, downgraded to 64 by 64 pixels, in the 193\,\AA\ channel, using the on-line 2-minute cadence synoptic data set at 1024 by 1024 pixels from 2010/05/01 through 2012/12/31 (\url{http://jsoc.stanford.edu/data/aia/synoptic/}). We chose the 193\,\AA\ channel because it reveals both flaring activity and the lower-energetic phenomena of eruptions from quiet-Sun regions. We then selected all events that exceeded a flare-like intensity and remapped their coordinates to an evolving synoptic map set, letting the signal fade over a 2-week period to allow patterns to stand out clearly. All pronounced clusters of events were subsequently reviewed in the daily summary movies of AIA observations, in the process eliminating instrumental artifacts related to spacecraft rolls and off-points, data gaps, and calibration mode data. 

As a next pre-selection criterion, we reviewed all M- and X-class flares in the AIA archive during the same period, and once again reviewed all daily summary  movies for those dates. We also collected candidate events during daily reviews of AIA data as annotated into the Heliophysics Events Registry \citep{aiadata}. From the sample of some five dozen candidate events, we selected the ten cases that were most compelling by visual inspection in supporting causal connections. 

For selected events, we also made and reviewed running-ratio movies in which the 211\,\AA, 193\,\AA, and 171\,\AA\ channels were combined. The frames in these image sequences were created by first computing logarithmic differences for time-averaged images (for 264\,s averages of fixed-exposure frames taken at 24\,s cadence and 2\,minute offsets between successive frames to be differenced for these movies), and then combining these in sets of three into the rgb color planes of a movie (clipping the scales at relative brightening or dimming to range from 0.1 to 10). These ``tri-ratio''image sets readily reveal changes in intensity, be they in bright active-region settings or in quiet-Sun or off-disk signals. The summary SDO/AIA data  for the selected events can be reviewed via the on-line Table$^{\href{http://www.lmsal.com/forecast/STYD.html}{S0}}$ at
\url{http://www.lmsal.com/forecast/STYD.html}.

For further review of the possible physical links between the selected data, we used potential-field source-surface (PFSS) models using the magnetogram assimilation code developed by \citet{schrijver+derosa2002b}, updated to assimilate SDO/HMI data (see \cite{schou+etal2011} for the instrument description), using the SolarSoft PFSS tool (see \url{http://www.lmsal.com/forecast/surfflux-model-v2/}. Links to visualizations of the field and an interactive viewing tool are included in the on-line table$^{\href{http://www.lmsal.com/forecast/STYD.html}{S0}}$.

In addition, we also compare the AIA observations with a magnetofrictional (MF) model developed by \citet{yeates+etal2008}, for which snapshots are included in the table. The MF model incorporates dynamics of the coronal field through the introduction of a dimensionless number that describes the balance between relaxation and diffusion and is the product of a friction coefficient and a diffusion coefficient \citep{mackay+vanballegooijen2006} . A second dimensionless parameter in the MF model is the ratio of the surface and coronal diffusion coefficients.  These two dimensionless parameters were originally calibrated by \citet{mackay+vanballegooijen2006} (a) such that flux ropes form above the internal polarity-inversion line of each active-region bipole with roughly one turn of twist and roughly once per 27 days, and (b) such that the field lines at the top of the computational box remained radial. The models shown in this study use the same parametrization as in the original model. We note that the MF model run requires selection of the magnitude and sign of the helicity for each emerging bipole; as these quantities are as a rule not known for active regions, these quantities are selected from a latitude-dependent distribution as described by \citet{yeates+etal2008}. Consequently, the MF model should be viewed as one possible incarnation of a dynamic corona, to be contrasted with the static PFSS model, but subject to major uncertainties related to the choice of the above mentioned parameters.

Other comparison material included the SOHO/LASCO data viewed using 
the JHelioViewer package (\url{http://www.jhelioviewer.org/}) and the STEREO browse data (with links in our on-line table, and directly accessible at \url{http://stereo-ssc.nascom.nasa.gov/browse}). 

In reviewing and selecting the events discussed in Section~\ref{sec:casestudies} we paid particular attention to the following possible mechanisms for sympathetic coupling: 1) (evolving) direct magnetic connection, 2) wave/front perturbation reaches distant region, and 3) distortion of overlying field by an eruption. We always allowed for mere synchronicity, and include discussion of such events in our case studies in the next section. While reviewing events, we generally tracked the duration of the glow of post-eruption arcades in the 171/193/211\,\AA-channel composite images; these durations can be found in the case studies denoted as $\Delta_{\rm A}$.

\begin{table*}
  \caption{Case studies: identified or suggested coupling mechanisms involving two or more activations within the solar corona. Listed are approximate starting times (UT) of the main events discussed in the text, followed by an indication of the regions involved. For 2010/08/01 and 2011/02/14-15 references are made to the cases discussed by Schrijver and Title (2011) and Schrijver et al. (2011). Where more than one active region number is listed, those following the first are abbreviated to two digits. Where specified, coordinates are given in arcseconds relative to disk center. Approximate times (UT) are given for the start of the initiating event.}
  \label{tab:couplings}
  \begin{center}
    \leavevmode
{%\tiny
    \begin{tabular}{llllll} \hline \hline              
 &  & \multicolumn{4}{c}{Likely primary coupling pathway} \\
  \cline{3-6} 
& & (evolving) direct & wave or propagating & distortion of overlying & (evolving) indirect \\
&  & connection         & disturbance           & field by eruption         & connection        \\
  \hline
A&2010/08/01 & 6:40 QS fil.~F$_1$, AR11092,-95,& - & 19:30 QS fil.~F$_3$&- \\
&           & connected with far-side regions&- &after fil.~F$_1$ erupt. &- \\
&&&&&\\
B&2011/02/14 &- & -&- & 18:00, 18:30, {\&} 19:00 AR11158, \\
&           &- & -&- & QS fil. at (-400,+450)\\
&&&&&\\
&2011/02/15 & 05:00 AR11158,-61& 1:45 QS fil. (-400,+450)&- &05:40 AR11158,-61 \\
&          &- &- &- & 07:00 {\&} 12:20 AR11158,-61,\\
&           &- &- &- & QS fil. (-400,+450)\\
&&&&&\\
C&2011/09/25 & 18:45 AR11295,11302,-03& 05:20 AR11303& -& 04:30 AR11301,-02\\
&&&&&\\
&2011/09/26 &21:00 AR11301,-02 &- &- &- \\
&&&&&\\
D&2011/11/09 &- &- &12:25 AR11341,-42 &16:30 AR11339,11342 \\
&&&&&\\
E&2011/11/22 &- &- &- &07:25 AR11353,-5,-7 \\
&           &- &- &- &10:30 AR11353,-4, polar crown\\
&&&&&\\
F&2011/11/30 &- &- &- &- \\
&&&&&\\
G&2011/12/11 &05:00 QS fil. to limb AR &- &- &- \\
&&&&&\\
H&2011/12/25 &-  &18:30 AR11385,-86 &00:15 two QS filaments & -\\
&          &-&-&separated by $\approx 5$\,h&-\\
&&&&&\\
I&2012/02/09 &17:00 (-300,+400), E-limb fil. &- &- &- \\
  \hline
 \end{tabular}
}
  \end{center}
\end{table*}
\section{Case studies}\label{sec:casestudies}
This section describes the events on a selection of 11 dates in detail. These dates are listed in Table~\ref{tab:couplings}, which also summarizes our conclusions regarding likely coupling mechanisms (identifying the times and the regions involved in each such set of events). On first reading, it may be easier first proceed to Section~\ref{sec:conclusions} in which we summarize the variety of processes before reviewing the following case studies in detail. An on-line Table$^{\href{http://www.lmsal.com/forecast/STYD.html}{S0}}$ provides access to the observational and modeling materials used the case studies.

\begin{figure}
\epsscale{1.18}\plotone{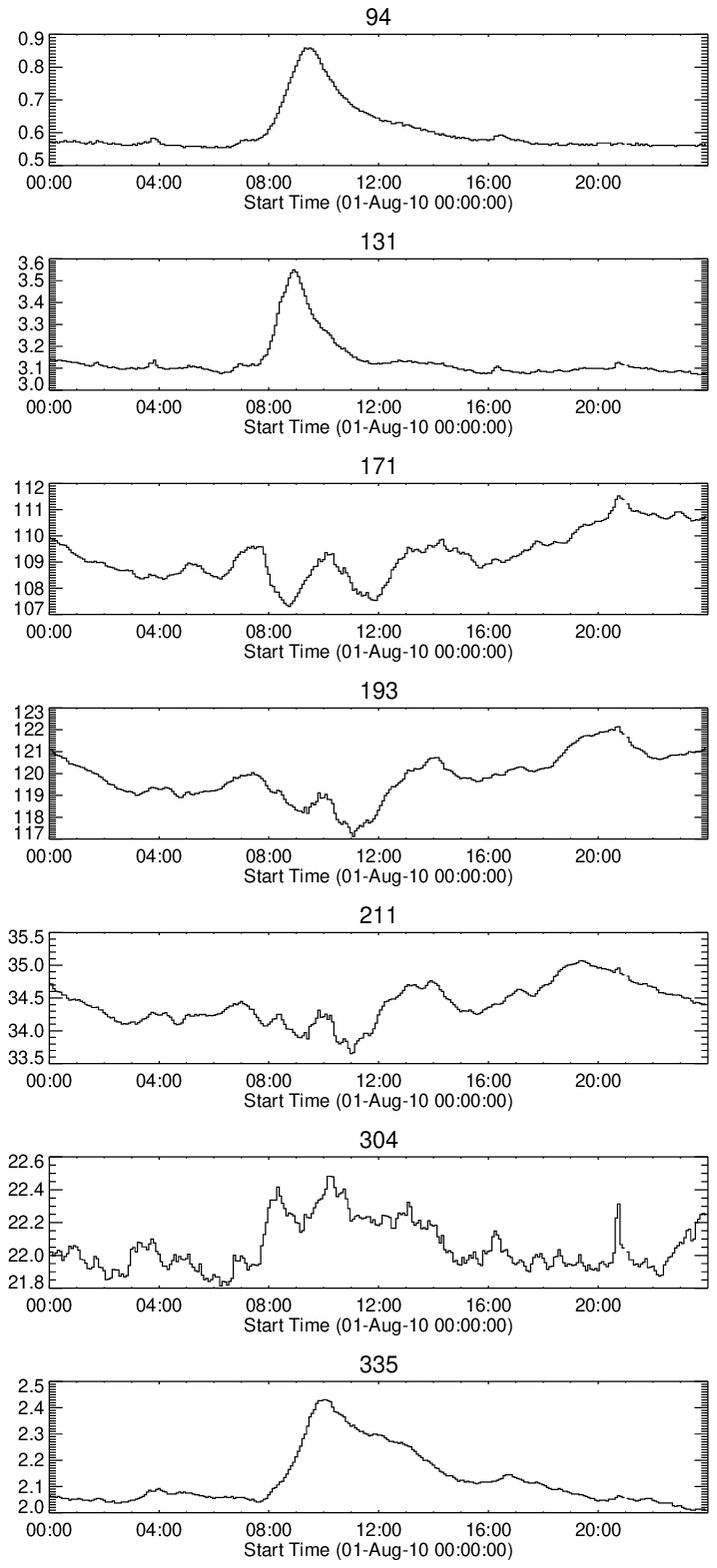}
\caption{AIA light curves for case A, 2010/08/01 (in DN/pixel/s).}\label{fig:lcurve20100801}
\end{figure}
\subsection{Case A: 2010/08/01}
%2010-Aug-1/2 :  5327/5328
The coronal events on 2010/08/01 have been extensively described by \citet{schrijver+title2010}. Following that study, the events in the solar corona and their impacts throughout the heliosphere have been studied in a series of papers that include the work by 
\citet{2012ApJ...750...45H}, % Harrison
\citet{2012ApJ...748...66M}, % Martinez
\citet{2011ApJ...739...43L}, %Li 
\citet{2012ApJ...746L..15L}, % Liu
\citet{2012ApJ...758...10M}, % Moestl
\citet{2012ApJ...749...57T}, % Temmer
\citet{2012ApJ...759...70T} - who describe the coronal field topology in detail - , % Titov
and \citet{2011JGRA..11612103W}. % Wu

\citet{schrijver+title2010} pointed out that the emergence of several new bipolar regions on the far side involved field configurations that directly connected all of the main sites of activity that day, involving a large-scale quasi-separatrix, connecting over several null points above the sites of filament destabilizations, and reaching into a flaring active region and a far-side CME. We refer to that paper for a detailed description of the events. 

\citet{schrijver+title2010}  argued that the magnetic topology was such that all of the major coronal events were connected by a quasi-separatrix involving several coronal null points. In an MHD modeling study of a simplified configuration, \citet{2011ApJ...739L..63T} demonstrated that the evolution of the common-envelope field over several filaments can cause flux ropes to erupt in the wake of a primary event. 

The evolution of the corona, visible in the AIA tri-ratio movies
$^{\href{http://www.lmsal.com/~schryver/STYD/AIAtriratio-211-193-171-2010-08-01T0000.mov}{S1},\href{http://www.lmsal.com/~schryver/STYD/AIAtriratio-211-193-171-2010-08-01T0000.mov}{S2},\href{http://www.lmsal.com/~schryver/STYD/AIAtriratio-211-193-171-2010-08-01T0000.mov}{S3},\href{http://www.lmsal.com/~schryver/STYD/AIAtriratio-211-193-171-2010-08-01T0000.mov}{S4}}$ as colored events, offers additional support for this process. In particular, the eruption of a large filament in the western hemisphere appears to induce large-scale reconnection that then allows a second filament eruption (starting around 06:40). The discoloration in the tri-ratio images, interpreted as a signature of coronal reconnection induced by the eruption, starts during the first, largest filament eruption. It is
most evident from about 2010/08/01 7:45\,UT through the end of that day ($\Delta_{\rm A}\ga 16$\,h). Fig.~\ref{fig:lcurve20100801} shows the AIA light curves for 2010/08/01, revealing (a) that the duration of the events differs considerably between the 94\,\AA, 131\,\AA, and 335\,\AA\ light curves, where there is one dominant peak with different decay time scales. Moreover, the 171\,\AA, 193\,\AA, and 211\,\AA\ light curves have a more complex structure that does not lend themselves to the interpretation as a single event. Hence, here (as in other cases below), we cannot rely on the disk-integrated light curves, but instead perform a visual inspection of the intensity and tri-ratio image sets to estimate the fading time scales in the emission patterns associated with the events discussed. 

The second main filament, lying to the south of and largely parallel to the first, erupts starting about 16:30\,UT with a noticeable rise speed and developing into
a rapid eruption by 19:30\,UT.
The delay of 9 to 12\,h between the initially observed thermal changes following the first filament eruption and the ultimate eruption of the second filament may appear rather long, but the thermal signatures of reconnection high in the corona in the wake of the preceding eruption of the largest filament after about 7\,UT continues until past the end of the day, so is clearly an ongoing process over many hours.

For these events, there is model support for causal linkages, both through the evolution of a series of PFSS models that illustrate the possible evolution of the field subject to flux emergence on the far hemisphere of the Sun as well as in a simplified MHD model of parallel flux ropes sharing a joint overlying field configuration. There are direct field connections between the destabilizing configurations, seen in the AIA images as in the field models. But despite all this indirect evidence of causal couplings, direct observational evidence is tenuous, leaving the ultimate interpretation rather subjective as to whether the overall field configuration is responsible for each event occurring with direct connections to the main topological feature, and as to whether the time of occurrence of the destabilizations is in fact affected by the evolution of the surrounding field. 

\begin{figure}
\epsscale{1.18}\plotone{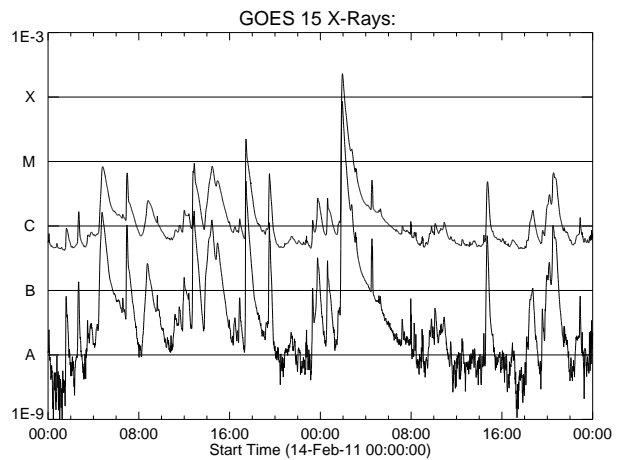}
\caption{GOES light curve for  case B, 2011/02/14-15.}\label{fig:goes20110214}
\end{figure}
\begin{figure}
\epsscale{1.18}\plotone{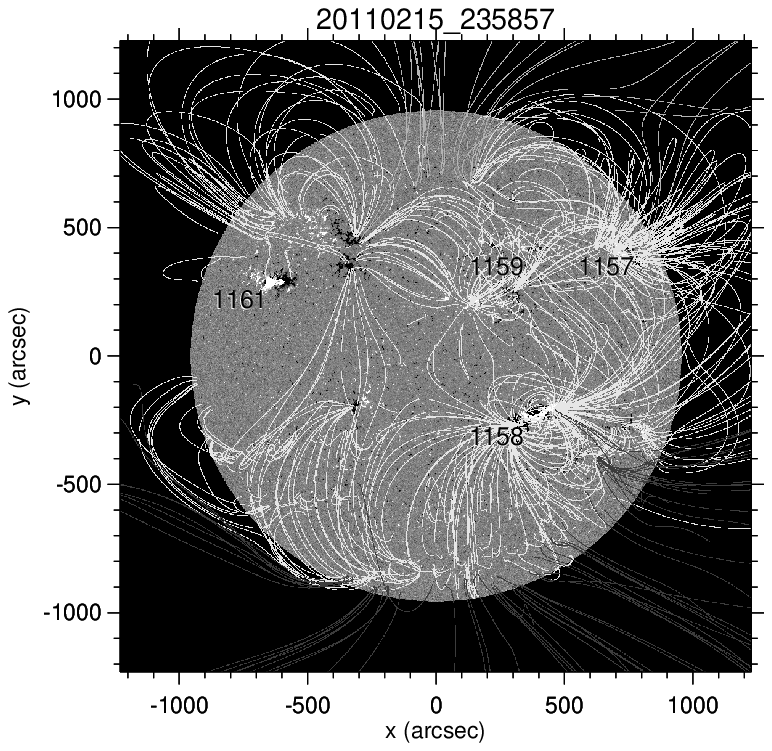}
\caption{HMI magnetogram, PFSS extrapolation, and AR numbers (showing the last four digits, as is a common standard, here and in subsequent similar figures for other cases) for  case B, 2011/02/15.}\label{fig:pfss20110215}
%\end{figure}
%\begin{figure}
\epsscale{1.18}\plotone{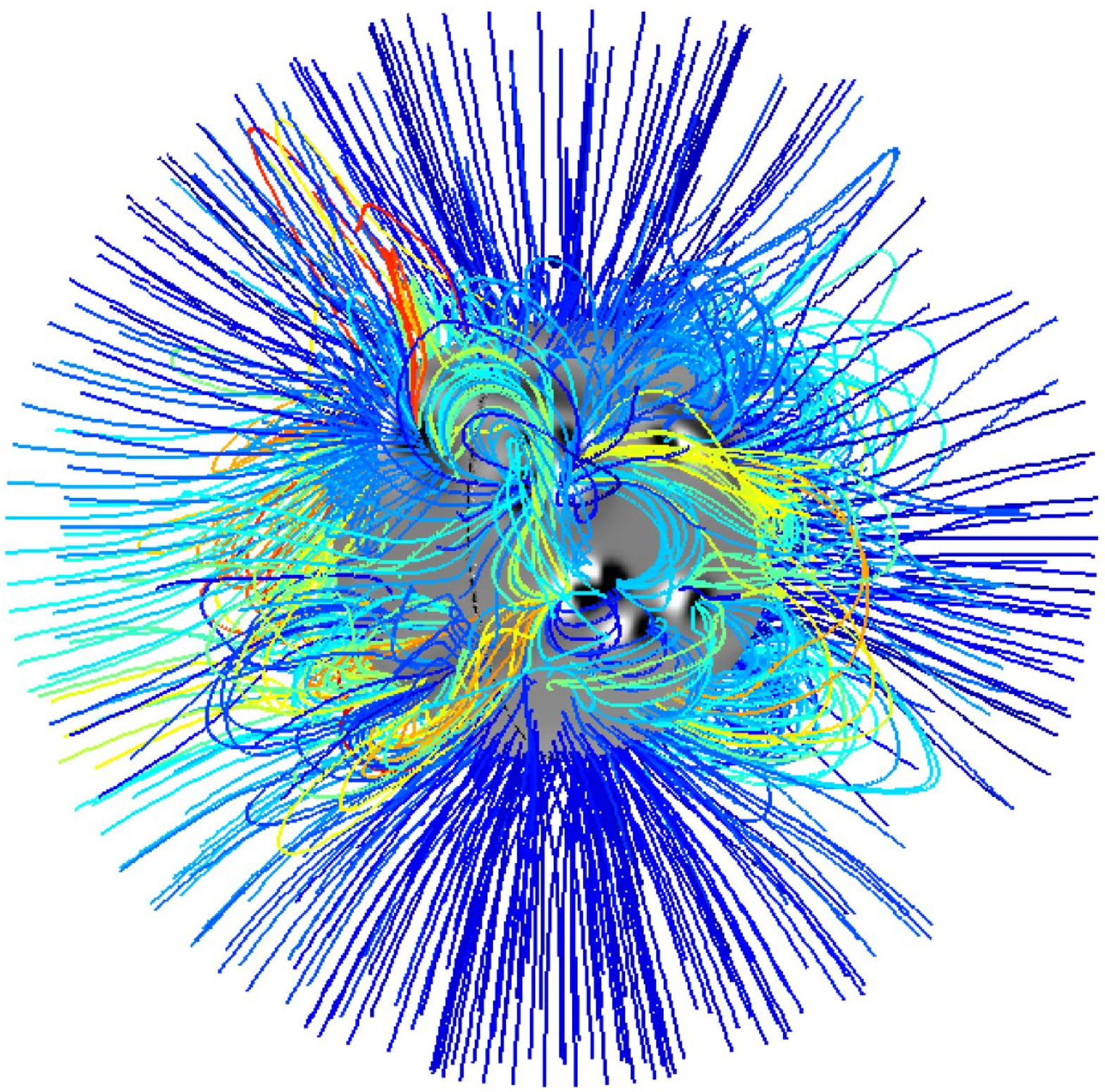}
\caption{Magnetofrictional field model for  case B for 2011/02/16, shown from the Earth perspective on the 14th. The field lines are colored depending on the angular spacing of their endpoints in latitude and longitude (regardless of whether they close on the surface or on the source surface) to reduce confusion in closely packed regions and in flux-rope configurations.}\label{fig:mf20110216}
\end{figure}
\begin{figure*}
\epsscale{1.18}\plotone{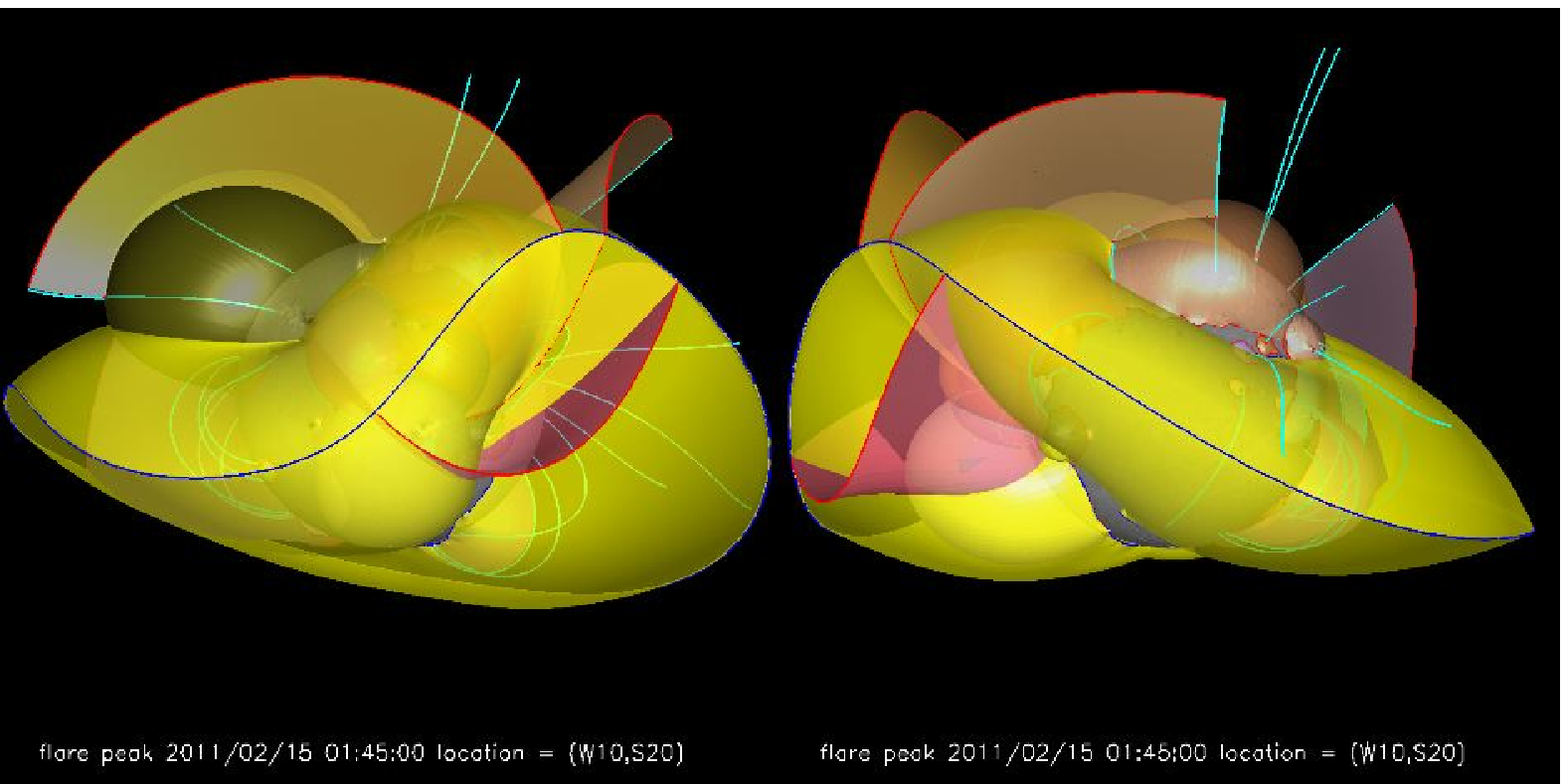}
\caption{PFSS field topology for  case B, 2011/02/15, from the perspectives of Earth and of STEREO-A. The main,  yellow, surface is the PFSS helmet, i.e., the envelope of the domain in the solar corona within which all field is closed. At its cusp the radial field component vanishes, so that this cusp forms the basis of the heliospheric current sheet above. Underneath and outside it are other surfaces that are the main topological domains defined by coronal null points (red dots), with their spine field lines shown in light blue. The dark blue curve is the null line for the radial field component on the source surface (at 2.5 solar radii). The red curves are the intersections of the domes and curtains with the solar surface and source surface, respectively.}\label{fig:topo20110215}
\end{figure*}
%2011-Feb-14/15 : 5524/5525
\subsection{Case B: 2011/02/14-15} 
The period around February 14 and 15, 2011, shows considerable activity on the Sun, with multiple flares and eruptions in the main region, AR\,11158, as well as other activity seen on disk and towards the northeast limb. 
The main center of activity is a substantial active region in the southern hemisphere, just past central meridian, AR\,11158. The region's X-class flare and associated CME are described in detail by, e.g., \citet{2011ApJ...738..167S} and \citet{2012ApJ...757..149S}. 
Here, we focus on February 14 and 15, when at least 20 C-class events occurred, one M2.2 flare, and the first X-class flare of sunspot cycle 23 (see the GOES light curve in Fig.~\ref{fig:goes20110214}).

We start our description with a high C-class flare around 2011/02/14 12:50UT. This event shows an eruption ($\Delta_{\rm A}\approx 20$\,m within the active-region core, with coronal signatures elsewhere last at least for 80\,m) from the trailing region of AR\,11158 towards the trailing part of AR\,11159 to its north, associated with a coronal propagating front that is most noticeable to the north of AR\,11158. This eruption occurs as a coronal reconfiguration is already in progress towards the north-east limb. The latter, although extending off the disk into the high corona, is not associated with substantial high-field activity in the SOHO/LASCO-C2 field of view. There is no obvious field connection to that region in either the PFSS or MF models on that day$^{\href{http://sdowww.lmsal.com/sdomedia/SunInTime/2011/02/14/f_HMImagpfss.jpg}{S10},\href{http://www.lmsal.com/~schryver/STYD/mfsnap5524.jpg}{S11}}$, but there is one in the model fields for a time five days later when more of the near-limb and over-the-limb field has been assimilated; we return to its discussion below. 

Around 2011/02/14 17:30\,UT, the same following part of AR\,11158 is involved in an M2.2 flare. This event is associated with a pronounced coronal propagating front, that is visible most prominently in all but the southerly directions, appearing to reach to the edges of the streamer belt towards the northeast, and out to the edges of three large topological domes converging on AR\,11159 (Fig.~\ref{fig:topo20110215}) (see also the on-line topology movie$^{\href{http://www.lmsal.com/~schryver/STYD/20110215topo.mov}{S17}}$). Associated with this flare is a dark 304\,\AA\ arc or surge from the trailing part of AR\,11158 moving towards the east and south, occurring around 18\,UT. At the same time, a small filament destabilizes, once again on this day, towards the northwest around $(-450,+500)$ (in arsceconds from disk center), in the same location as the field reconfiguration noted for the events following 12:50\,UT, and within 20\,minutes of the propagating front reaching a decayed active region just to the southwest of it near $(-400,+450)$ .

Then, around 18:30\,UT, there is another eruption from the same part of AR\,11158, again associated with activity over the filament configuration to the NE. This happens once more just after 19\,UT (associated with another dark 304\,\AA\ feature moving westward) in events that are then followed by a high C-class flare around 19:30\,UT ($\Delta_{\rm A}\approx 30$\,m within the region and about 2\,h in its immediate vicinity). In the trailing end of that event, another dark 304\,\AA\ feature is seen, forming threads that reach a long way towards AR\,11161.

The X2.2 flare occurs at 2011-02-15 starting at 01:45\,UT ($\Delta_{\rm A}\approx 60$\,m within the region and about 90\,m outside it).  This flare is associated with a pronounced coronal propagating front and CME, studied in detail by  \citet{2011ApJ...738..167S}. The filament region to the northeast near $(-450,+500)$ again shows limited synchronous activity. This event set appears coupled through the propagating disturbance, which may be either a pure wave or an expansion front that couples to the filament in the northeast through higher field breached by the initiating CME. 

In the decay phase of this main flare (brightest within the central regions of AR\,11158) another eruption from the trailing segment or the active region occurs around 04:30\,UT ($\Delta_{\rm A}\approx 5$\,m in the active region core, some 90\,m in its far surroundings). This time, the 304\,\AA\ channel shows a moving, dark connection from AR\,11158 to AR\,11161 towards the northeast limb. This field deformation is a precursor to a mid-C class flare from that region around 05\,UT, that shows a similar, subsequent dark structure in 304\,\AA\ continuing its motion towards the east and south, now with synchronous brightening and deformations in AR\,11161, around 05:40\,UT being highly suggestive of a direct connection between the region, but falling short of showing convincing evidence, even in high-contrast running-difference image sequences (not shown). By 07\,UT, this pair of regions again shows synchronous mild flaring and surging, now involving also a ribbon-like brightening, at least in 304\,\AA, in the quiet-Sun small filament towards the northeast limb near $(-450,+500)$ where repeated activity occurred as described above. 

Later in the event sequence, AR\,11161 and the filament configuration around $(-450,+500)$ show pronounced synchronous activity once 2011/02/15 12:20\,UT, with some activity also at $(+450,+500)$ in the compact bipole north of AR\,11159. 

In between the several near-synchronous events pointed out above, each of the regions involved exhibits activity (both flares and eruptions) in the absence of synchronous activity in the other regions, including, for example, the ultimate eruption of the filament towards the northeast limb after about 23:20\,UT.

The general geometry of the corona on the Earth-facing side of the Sun$^{\href{http://sdowww.lmsal.com/sdomedia/SunInTime/2011/02/15/f0171pfss.jpg}{S19}}$ appears to be described quite well by the PFSS configuration shown in Fig.~\ref{fig:pfss20110215} (showing the configuration one day later when AR\,11161 that emerged on the far side had been assimilated into the model). The magnetofrictional model for that day$^{\href{http://www.lmsal.com/~schryver/STYD/mfsnap5524.jpg}{S16}}$ shows a connectivity similar to that of the PFSS model, suggesting that there are no pronounced flux-rope configurations for 2011/02/14. We return to this below, however, arguing that this may be because that formation is delayed in the MF model relative to reality. 

The model field and its topological summary in Figs.~\ref{fig:pfss20110215} and~\ref{fig:topo20110215} show (a) a null point floating above about $(-100,-100)$, 
and (b) a helmet structure that arches over AR\,11158 connecting to the surface
northeastward of AR\,11161. The repeated eruptions in the field connecting AR\,11158 and AR\,11159 arch towards the coronal null, and must somehow connect through it if it is to explain the synchronous activity in AR\,11161. Any reconnection occurring through the coronal null requires corresponding changes on both sides, thus providing a pathway to explain some of the sympathetic activity in AR\,11158 and the field  of and near AR\,11161 in terms of coupled fields, and thus in terms of sympathy.  
Eruptions that become CMEs have to breach the field under the helmet configuration, which involves field reaching down to the AR\,11169 and its surroundings (as can be seen in the left panel of Fig.~\ref{fig:topo20110215}). 

In view of the repeated near-synchronous activity and suggestions of connecting signals (surges or waves), one might conclude that the dynamics of the evolving field forced by the frequent activity in AR\,11158 is likely not adequately captured by the approximations of a static (PFSS) or slowly evolving (MF) field. We also note that the details of the  high configuration of the field for these days is quite sensitive to changes in the solar surface field. The PFSS field based on the revised assimilation procedure (``Version 2'', see \url{http://www.lmsal.com/forecast/surfflux-model-v2/}) shown in this study and that computed for Version~1 in \citet{2011ApJ...738..167S} show different configurations for the helmet streamer than those in Fig.~\ref{fig:topo20110215}. On the other hand, most of the solar surface for this date appears to lie under the warped, undulating helmet configuration, so invoking the disruption of that configuration by nascent CMEs in causal linkages would not be readily falsifiable in this case. Moreover, the many non-synchronous events in the main activity centers demonstrates that care needs to be taken in interpreting the synchronous ones as consequences of sympathetic coupling even if pathways in the magnetic field are identified.

In the end, we propose that the events on 2011/02/14-15 are, in fact, causally connected. This is based on a view of the MF configuration for 2011/02/16, when more near-limb field has been assimilated into the models. This configuration in shown in Fig.~\ref{fig:mf20110216} shown from the Earth's perspective for 2011/02/14. This field model clearly shows a rope connecting AR\,11158 to AR\,11161, as it links through AR\,11159, and it shows another high rope-like configuration (shown in red) from AR\,11161 to above the location from where the northernmost quiet-Sun filament lifted off. This model connection is supported by the repeated activation of the filament towards the northeast limb closely following activity in AR\,11158. The southern base of that rope connects to a region of major flaring in the trailing part of AR\,11158, where another rope goes southeastward into the direction of the major surge around 2011/02/15 04\,UT. Here, what appear to be ropes connect all of the regions of activity involved, and the footpoints of all ropes are directly adjacent from one connecting segment to the next, if not in fact interlinked.

In conclusion, we find evidence of synchronous activity in directly connected regions where the connecting field is affected by flares or eruptions in at least one of these connected regions. The synchronicity of events in more distant regions, both in quiet-Sun field and to active-region field, may be by chance, but the MF model for the configuration a few days after the above events (when more of the field involved is on disk and assimilated into the models) is highly suggestive of direct connections between all regions involved.

\begin{figure*}
\epsscale{1.18}\plotone{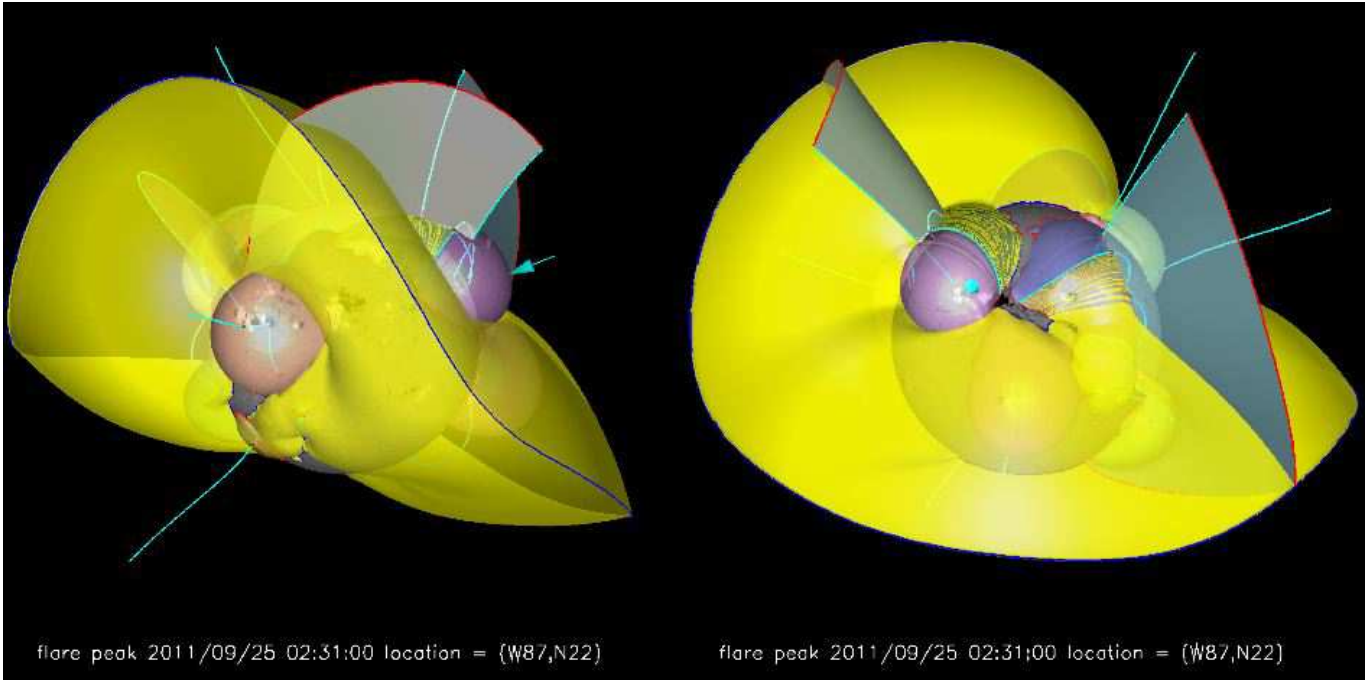}
\caption{PFSS field topology for  case C, 2011/09/25 (cf. Fig.~\ref{fig:topo20110215} for a description of the details).}\label{fig:topo20110925}
\end{figure*}
\begin{figure}
\epsscale{1.18}\plotone{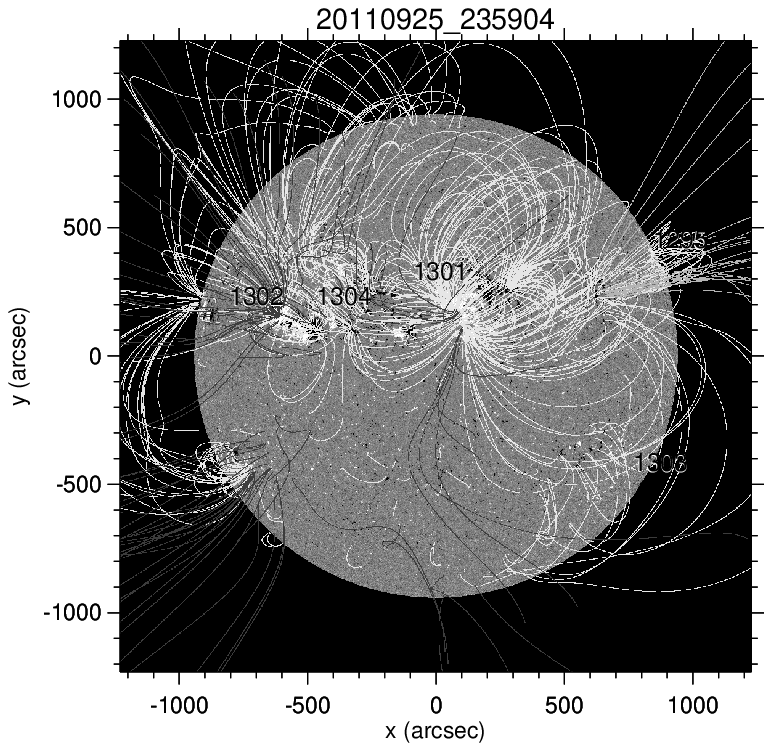}
\caption{HMI magnetogram, PFSS extrapolation, and AR numbers for case C,  2011/09/25.}\label{fig:pfss20110925}
%\end{figure}
%\begin{figure}
\epsscale{1.18}\plotone{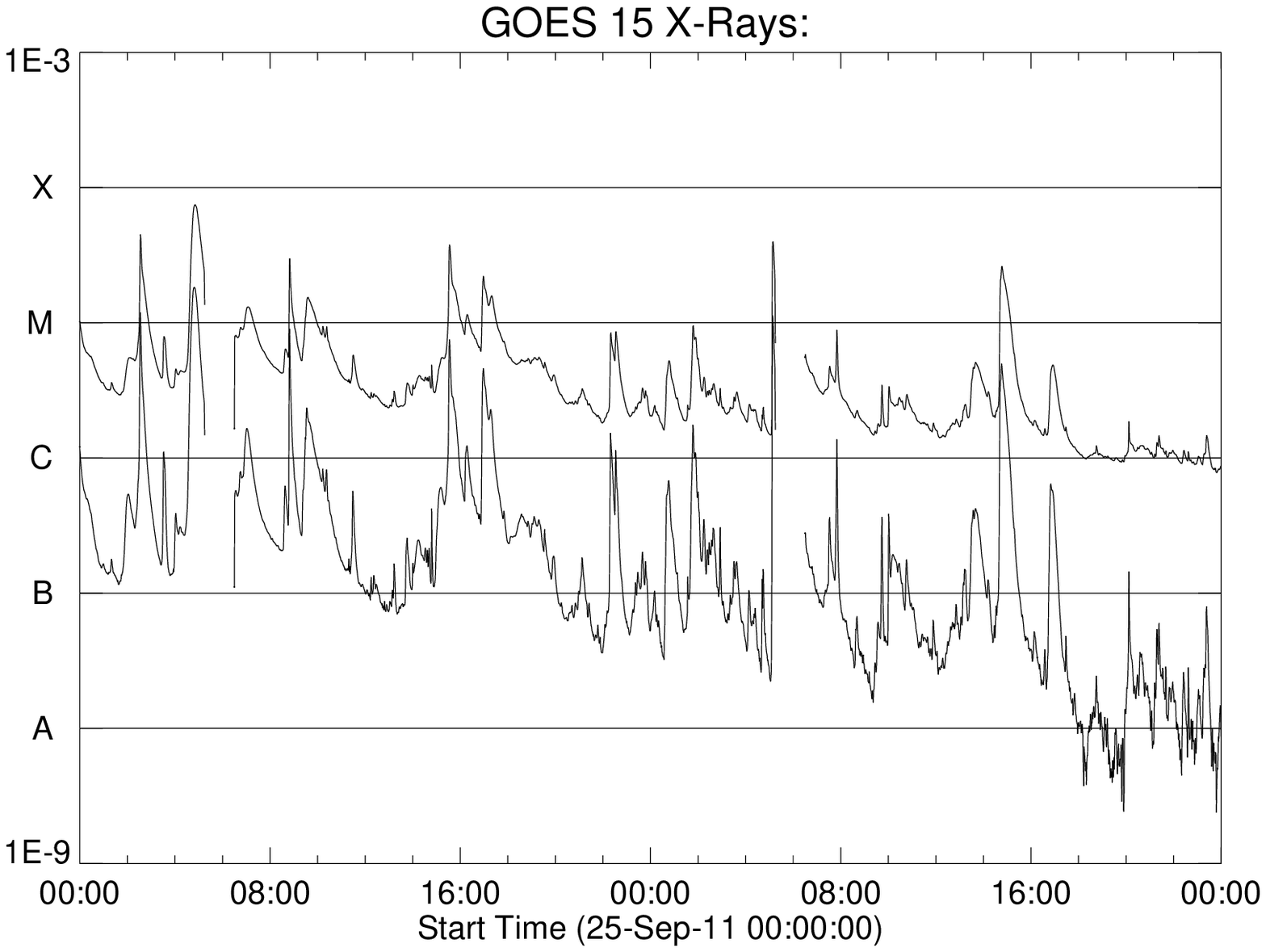}
\caption{GOES light curve for  case C, 2011/09/25.}\label{fig:goes20110925}
\end{figure}
\begin{figure}
\epsscale{1.18}\plotone{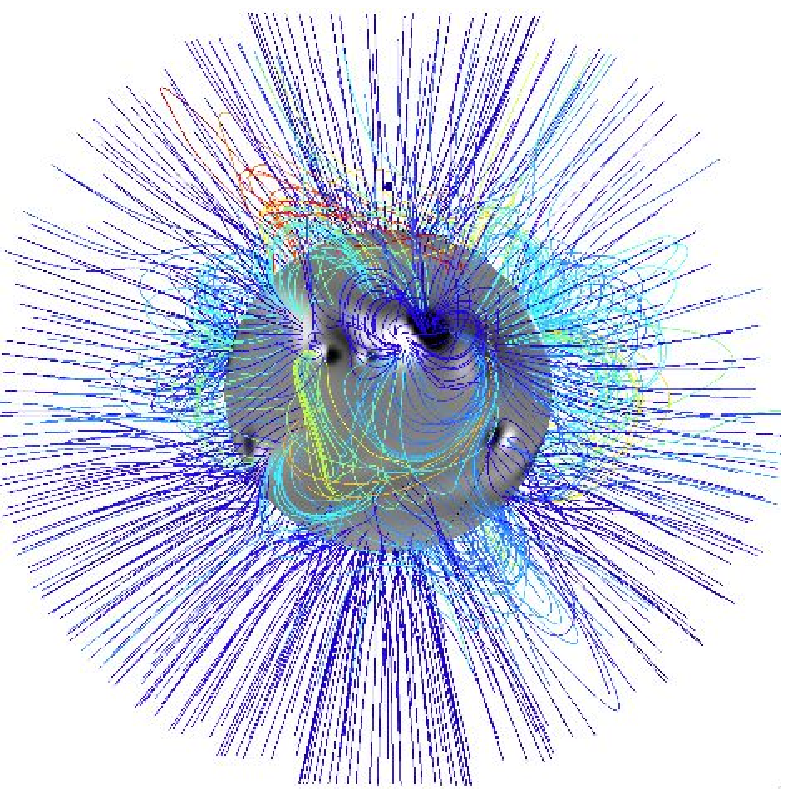}
\caption{Magnetofrictional field model for  case C, 2011/09/25 (cf. Fig.~\ref{fig:mf20110216} for a description of the details).}\label{fig:mf20110925}
\end{figure}
%2011-Sep-25/26 : 5747/5748
\subsection{Case C: 2011/09/25-26} 
The final week of September of 2011 showed considerable activity:
along with a multitude of lesser events, 
there were 8 M flares on the 24th, 6 on the 25th, and 2 more on the 26th. Here, we focus on 
the activity on 2011/09/25 and 26 when the main regions were well onto the disk and the overall magnetic patterns readily observable and models subject to less uncertainty. 

The 25th begins in the decay phase of an eruptive M1.0 flare from AR\,11303 in the southwest (see Fig.~\ref{fig:pfss20110925}), associated with a CME as seen in SOHO/LASCO. This is followed, around 01:15\,UT, by near-synchronous eruptions from the northwest limb and from the trailing side of AR\,11302. These events occur on opposite sides of the helmet streamer and are unlikely to be physically connected (cf. Fig.~\ref{fig:topo20110925}). 
 
The first large flare, class M.4., starts in the GOES data at 02:27UT in AR\,11302 ($\Delta_{\rm A}\approx 5$\,m), in the northern hemisphere and east of central meridian. At about 02:35\,UT, the flare reaches its peak in the AIA channels (with a very pronounced diffraction pattern revealing a compact kernel). At that time AR\,11303 at the southwest limb is also brightening, reaching its peak brightness (also with a pronounced diffraction pattern) just minutes later. The diffraction patterns in these two events, indicative of compact flare kernels, brighten within about 70\,s of each other; for a signal traveling between these two regions over the solar surface, this would require velocities in excess of 20,000\,km/s. The synchronicity is thus either associated with energetic particles, occurs simply by chance, or there is some other evolution that causes both of these events to go off simultaneously. We find no unambiguous evidence for such an intermediary event in either the observations or in the model fields, but do point out that there was a preceding CME from the direction of AR\,11303 visible in SOHO/LASCO C2 from 00:36\,UT onward and some 10\,min.\ earlier in STEREO/A COR1; that CME propagates in fairly narrow cone towards the southwest, with no obvious signatures over the northeast limb above AR\,11302.

The GOES light curves (Fig.~\ref{fig:goes20110925}) show signatures of a double flare, being dominated by the signal from AR\,11302. The flare from AR\,11303 is associated with a mild coronal propagating front traveling over some tens of degrees, and an eruption into the overlying high corona. The flare/eruption from AR\,11303 develops into a CME over the southwest limb. We note that for this pair of distant events, the field models in Figs.~\ref{fig:mf20110925} and~\ref{fig:topo20110925} (see also the on-line topology movie$^{\href{http://www.lmsal.com/~schryver/STYD/20110925topo.mov}{S18}}$) do not reveal connections between these regions (with the PFSS model putting ARs\,11302 and\,11303 under disjoint domains of connectivity), nor do running-ratio movies reveal any wavelike perturbation connecting these events that stands out above the noise.
 
At 2011/09/25 04:31\,UT an eruptive M7.4 flare initiates from AR\,11302, with a pronounced coronal propagating front moving in southerly directions ($\Delta_{\rm A}\ga 3$\,h for the coronal region towards the southwest from the active region, still evolving when an Earth transit starts at 06\,UT)  that develops into a CME from the Sun in a southern to southeasterly direction. Apart from a synchronous brightening of the core of AR\,11303 and compact, weak brightenings in quiet Sun in the region leading AR\,11301, there is no other obvious synchronous activity elsewhere. But AR\,11303 does brighten once more, starting around 05:20\,UT, around the time that the perturbation front from the eruption of AR\,11302 approaches it.

There is a CME visible in SOHO/LASCO C2 from 8\,UT onward, most strongly from the direction of AR\,11303; no AIA observations of its origin are available owing to an Earth transit from 06:05\,UT to 07:20\,UT.

The M3.1 flare starting at 2011/09/25 08:46UT in AR\,11302  ($\Delta_{\rm A}\approx 5$\,m) is compact once more with no clear sign of any high coronal field being breached or distorted. 

Shortly after 09\,UT an eruption begins from AR\,11303, clearly opening field starting around 09:35\,UT, coincident with activity at the top of a polar-crown prominence in the southwest. This eruption is associated with an M1.5 flare starting in the GOES data at 09:25\,UT. Around 09:30\,UT there is a compact brightening in the core of AR\,11302. Again, there is synchronicity between ARs\,11302 and~11303 without model or observational support for a causal connection. At other times, however, such as the flare in AR\,11302 around 10:10\,UT, no synchronicity between events in these regions is observed.

At 15:26\,UT a pronounced southward eruption (and surge/spray) is associated with an M3.7 from AR\,11302  ($\Delta_{\rm A}\approx 30$\,m in the region, and until past the next flare 90\,m later in its surroundings), associated with a wide-angle CME from the southeast to the north. This is followed by a compact brightening into an M2.2 flare starting at 16:51UT in the same region.

Around 2011/09/25 18:45\,UT, an eruption starts over the northeast limb over regions trailing AR\,11302  (with a rather subjective estimate of $\Delta_{\rm A}\approx 6$\,h). Once that eruption accelerates around 19\,UT, a high loop configuration over the western equatorial limb, connecting regions near AR\,11303 and AR\,11295, also begins to rise. Around 21\,UT, high loops connecting ARs\,11301, 11302, and 11304 expand  ($\Delta_{\rm A}\approx 3$\,h), as the western-equatorial loop system continues to expand,followed by repeated mid-C level flaring in AR\,11303 with an eruption around 2011/09/26 02:20\,UT. Field models do not suggest how any causal linkage between these events might occur, except for the direct linkage of ARs\,11301 and~11302.

Then, there is another compact brightening in AR\,11302 
associated with an M4.0 flare starting at 2011/09/26 05:06UT  ($\Delta_{\rm A}\ga 1$\,h with reconfiguration of the surroundings, but observations are cut short by an Earth transit), and finally a moderately disruptive M2.6 starting at 2011/09/26 14:37UT  ($\Delta_{\rm A}\approx 90$\,m). Both events are limited to activity within AR\,11302.

The overall topology of the coronal field, as summarized in Fig.~\ref{fig:topo20110925}, does not support connections between the (near-)synchronous events in ARs\,11302, 11303, and 11295. These regions appear isolated by separatrices: ARs\,11302 and 11295 lie on either side of the tilted (yellow) separatrix surface (the ``helmet'') seen in the PFSS model, while AR\,11303 lies under that helmet. Hence, the eruption of one region is not connected to either of the others through topological structures. On the other hand, the reconfiguration of the loops seen between ARs\,11302 and 11301 are directly mappable to the magnetic field, most clearly seen in the MF model in Fig.~\ref{fig:mf20110925}. These are not, however, associated with major flaring. We conclude that in this case, despite several instances of synchronicity, there is no compelling observational evidence for causal connections between the main events on 2011/09/25 and 26 except, perhaps, for the coupling through a field disturbance (propagating front) around 05:20\,UT on 2011/09/25.

\begin{figure}
\epsscale{1.18}\plotone{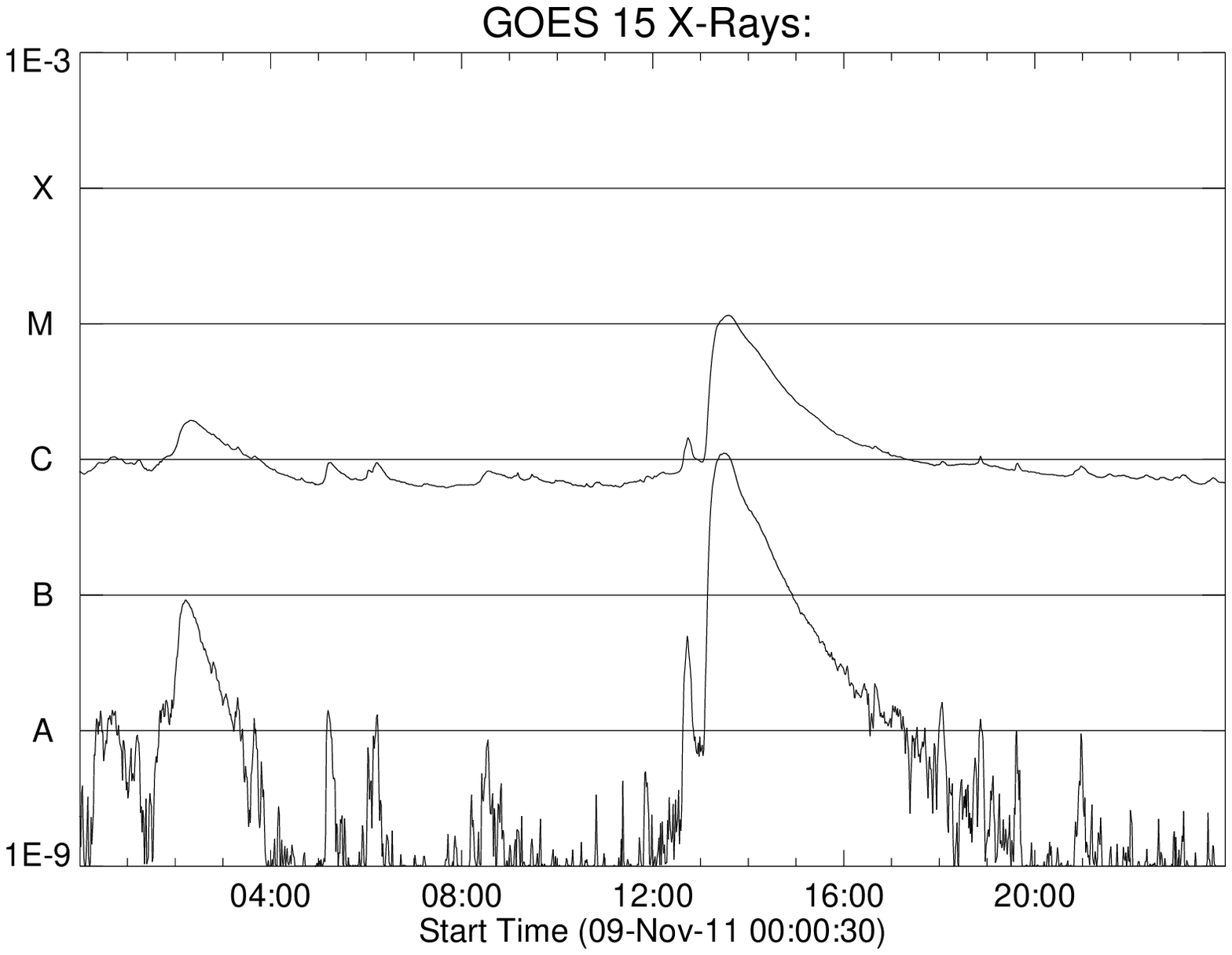}
\caption{GOES light curve for  case D, 2011/11/09.}\label{fig:goes20111109}
%\end{figure}
%\begin{figure}
\epsscale{1.18}\plotone{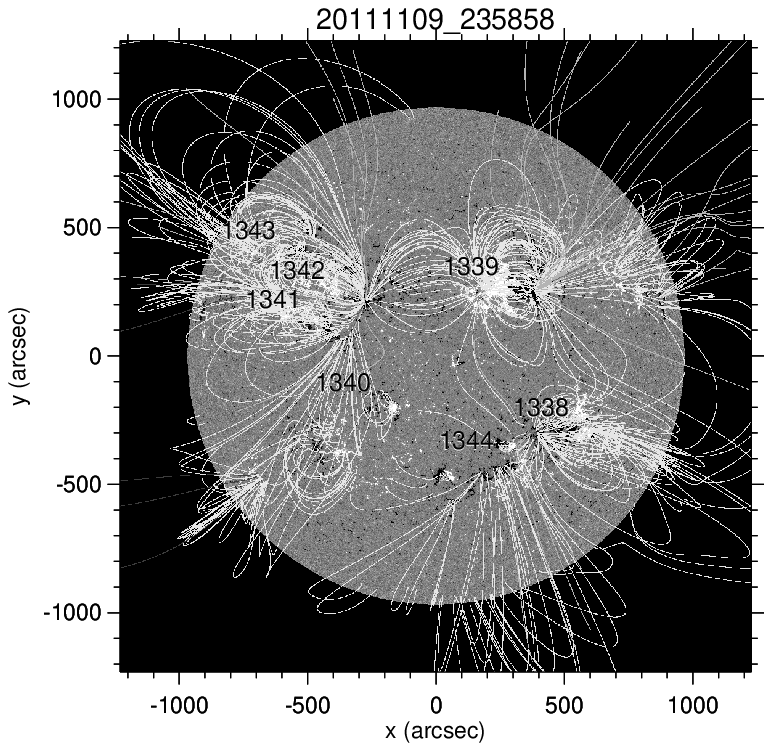}
\caption{HMI magnetogram, PFSS extrapolation, and AR numbers for case D,  2011/11/09.}\label{fig:pfss20111109}
\end{figure}
\begin{figure}
%\epsscale{1.18}\plotone{mf20111109_ropes.eps}
\epsscale{1.18}\plotone{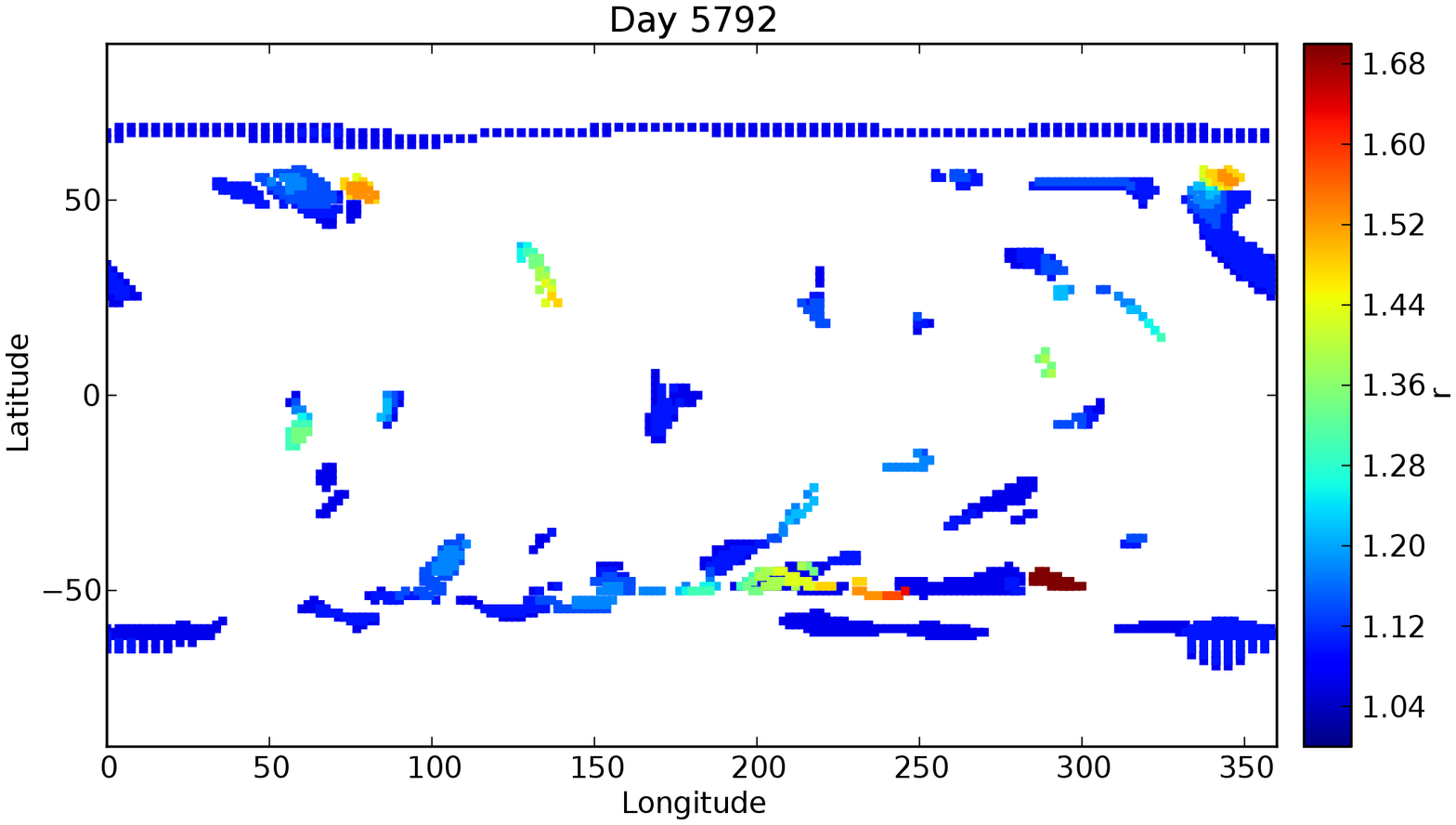}
\caption{Flux rope positions, colored by height (see color bar, in units of solar radii), from the magnetofrictional model for case D. The Carrington longitude of disk center on 2011/11/09 is 96$^\circ$.}\label{fig:mf20111109_ropes}
%\end{figure}
%\begin{figure}
\epsscale{1.18}\plotone{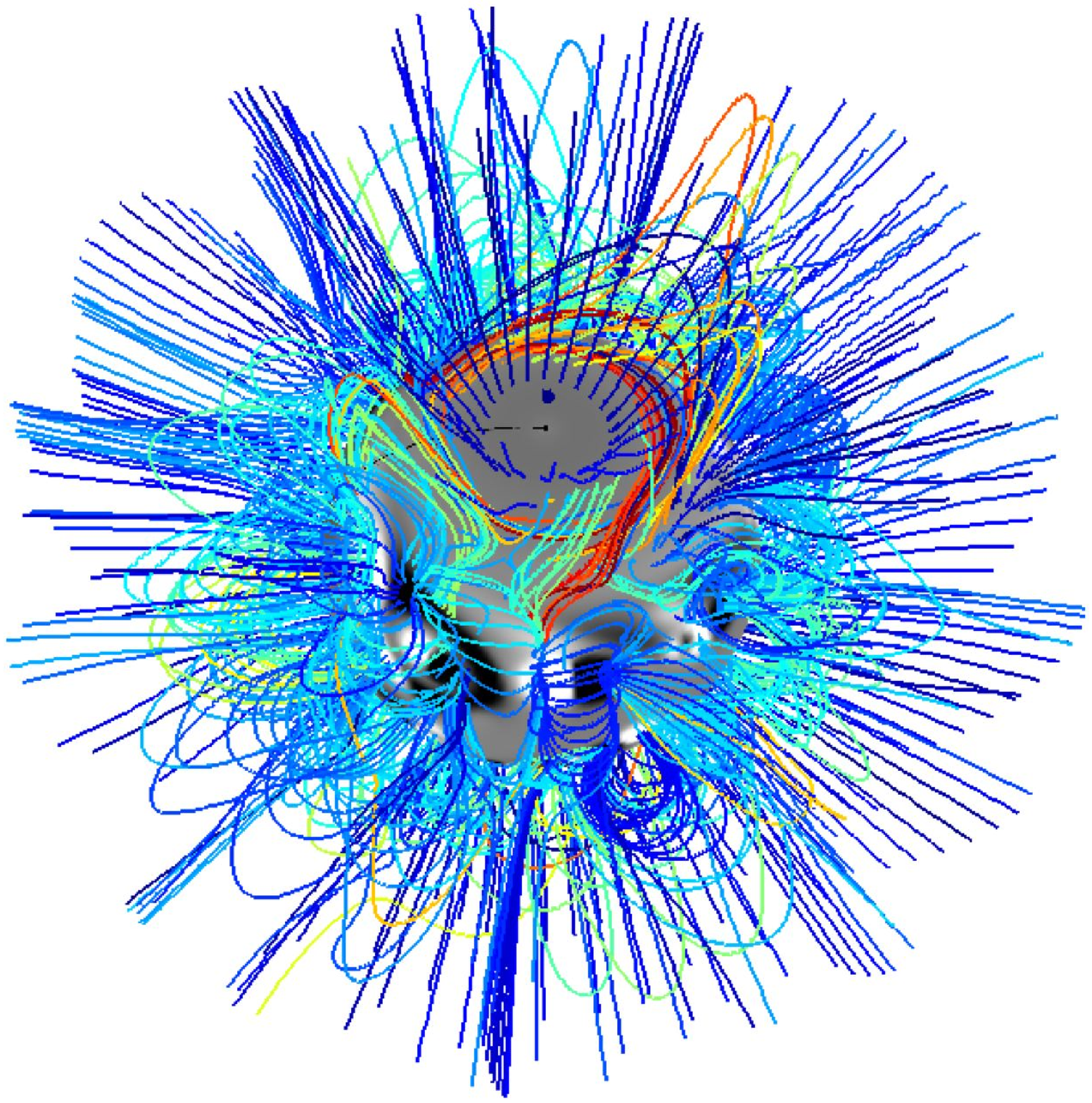}
\caption{Magnetofrictional field model, tipped to a viewing angle of 57 degrees north latitude for  case D, 2011/11/09  (cf. Fig.~\ref{fig:mf20110216} for a description of the details).}\label{fig:mf20111109_tip57}
\end{figure}
%2011-Nov-9 :  5792
\subsection{Case D: 2011/11/09}
Around 07:30\,UT, a large quiet-sun filament configuration erupts in the southeastern quadrant; the post-eruption arcade continues to gradually evolve throughout the remainder of the day. Around 11:30\,UT, an eruption initiates behind the northwestern limb, with effects in the visible part of the corona fading out around 12:30\,UT. Neither of these events appear to be linked, either in the AIA observations or  in the field models, with the major flaring that follows. 

At about 12:25\,UT an M1.1 flare begins (flagged as 13:04 in GOES data), initiated in AR\,11341, associated with a large filament eruption from neighboring AR\,11342 that comes to full development around 13:00\,UT, resulting in a full-blown CME  ($\Delta_{\rm A}\approx 8$\,h). The rising eruption distorts and disrupts overlying field, with signatures seen (in, e.g., the tri-ratio movie$^{\href{http://www.lmsal.com/~schryver/STYD/AIAtriratio-211-193-171-2011-11-09T0000.mov}{S5},\href{http://www.lmsal.com/~schryver/STYD/AIAtriratio-211-193-171-2011-11-09T0600.mov}{S6}}$), for example, into the inter-regional separatrix in the trailing polarity of AR\,11339 to the west of the erupting region that are visible until past 14\,UT, with the post-eruption arcade itself glowing until at least 20\,UT.

After the thermal signatures of the previous eruption over the surrounding quiet Sun have faded away, by about 16:30\,UT, there are coronal deformations to the north of AR\,11339 just south of the polar coronal hole boundary. These appear to be caused by an eruption  ($\Delta_{\rm A}\approx 2-6$\,h, poorly determined because of the slow evolution in the end), of which we see faint ribbon-like signatures northwest of AR\,11339 starting around 17:45\,UT, as well as high off-limb field deformation towards the north from there, with tilting field reaching from there to about 45 degrees clockwise from the north.  SOHO/LASCO and STEREO/SECCHI show faint blowouts associated with this event. When the 94\,\AA\ channel is suggestive of high-temperature loops forming after that eruption, there is simultaneous formation of such hot loops within an evolving environment in the region adjacent to AR\,11342 towards the northwest, both lasting until at least 21\,UT. 

There are high coronal flux ropes connecting these features between 1.3 and 1.7$R_\odot$ in the magnetofrictional model (Fig.~\ref{fig:mf20111109_ropes}, in particular at 50$^\circ$ north over AR\,11342 and surroundings). Fig.~\ref{fig:mf20111109_tip57} shows a tilted perspective of the magnetofrictional field model, revealing that the erupting AR\,11341 and the eruption north of AR\,11339 have connections that end adjacent to each other (close to the central meridian at the center of the disk in this tilted perspective) showing that there are likely indirect magnetic interactions between the evolution of these structures, but not demonstrating their involvement in sympathy.

\begin{figure}
\epsscale{1.18}\plotone{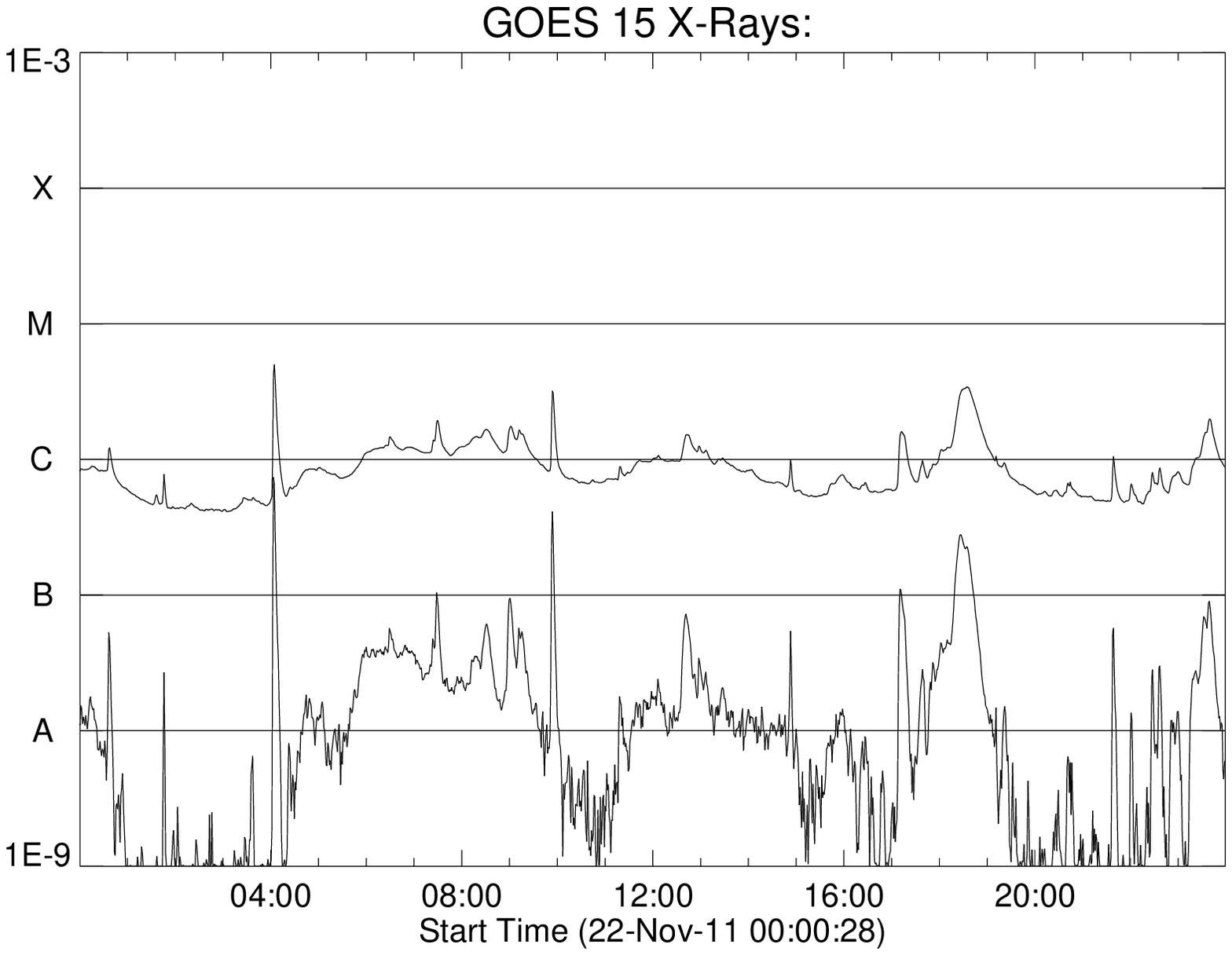}
\caption{GOES light curve for  case E, 2011/11/22.}\label{fig:goes20111122}
%\end{figure}
%\begin{figure}
\epsscale{1.18}\plotone{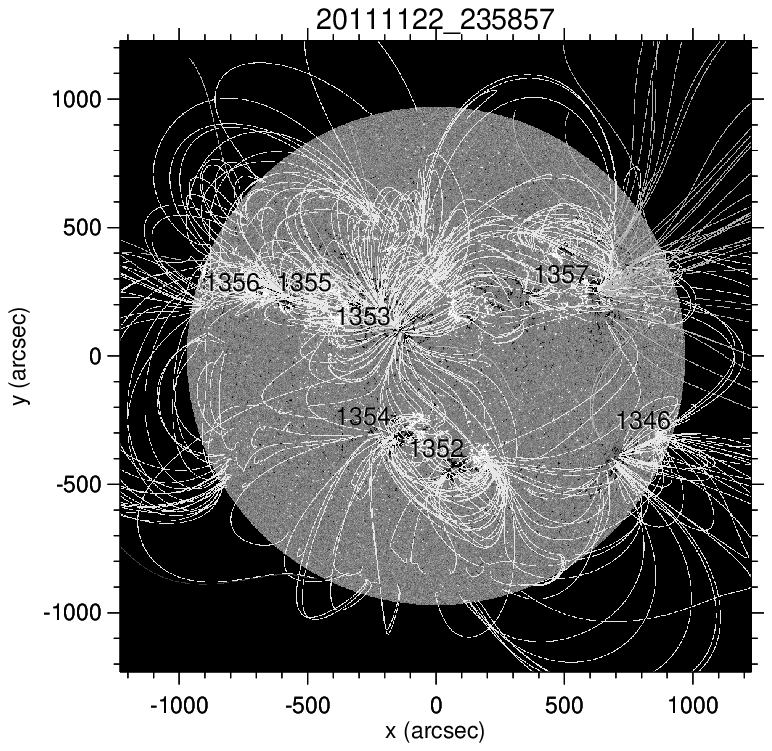}
\caption{HMI magnetogram, PFSS extrapolation, and AR numbers for case E,  2011/11/22.}\label{fig:pfss20111122}
\end{figure}
\begin{figure}
\epsscale{1.18}\plotone{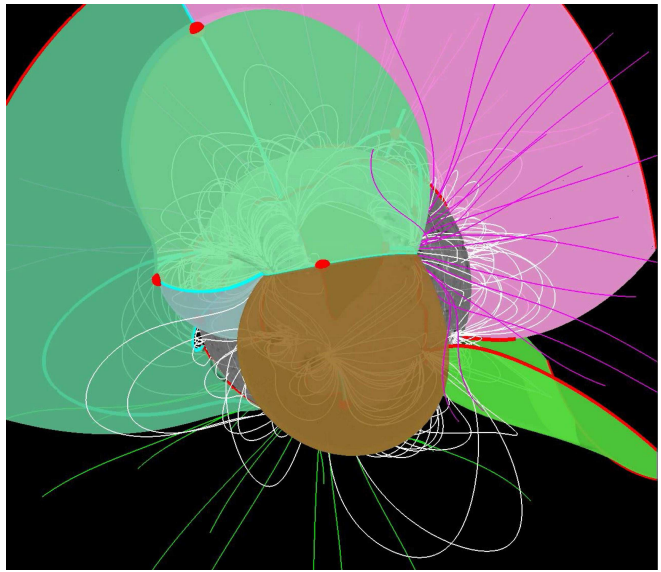}
\caption{Partial rendering of the topological domains for the PFSS model for  case E, 2011/11/22. Selected domain boundaries are shown as translucent surfaces for those separatrices associated with coronal null points (shown as red balls on the blue spine field lines).}\label{fig:topo20111122}
\end{figure}
%2011-Nov-22 : 5805
\subsection{Case E: 2011/11/22}
Shortly after 2011/11/22 04\,UT, a C7 flare goes off in a compact emerging-flux region slightly to the south of the line connecting ARs\,11355 and\,11356 (see Fig.~\ref{fig:pfss20111122}). Just as that event ends, a small activation and eruption (and CME observed in both SOHO/LASCO C2 and STEREO/SECCHI-A COR1) occurs to the north of it  ($\Delta_{\rm A}\approx 10$\,h), in the leading edge of AR\,11356, expanding somewhat to the northeast and southwest through about 07\,UT. There are synchronous brightenings visible at the trailing end of the polar-crown configuration north of AR\,11353, which itself shows a high coronal reconfiguration starting at about 07:25\,UT, when there are simultaneous brightenings south of AR\,11355 and in AR\,11357 far to its west. These events are associated with a CME first seen in SOHO/LASCO C2 after 06:24\,UT. All of these regions lie under a large dome of connectivity in the PFSS model, mostly near its periphery, and thus suggest involvement of field that connects through a coronal domain near a high coronal null (see Fig.~\ref{fig:topo20111122}).

Around 10:30\,UT, the filament configuration connecting AR\,11353 in the northern hemisphere and AR\,11354 in the southern hemisphere activates  ($\Delta_{\rm A}\approx 4$\,h), synchronous with brightenings and loop shifts from there to the northern polar crown, including the corona to the north of, and apparently high over, AR\,11353. Around 13:50\,UT, there are simultaneous brightenings reaching from AR\,11356 to near that same position in the northern polar crown configuration. Here, Fig.~\ref{fig:topo20111122} suggests the connections to involve two adjacent domains connected at a coronal null westward of  AR\,11353.

At around 17:10\,UT there is another, more gradual mid-C class brightening in the emerging flux region southeast of AR\,11355  ($\Delta_{\rm A}\approx 100$\,m), not obviously associated with any other substantial activity.  Then, just past 18\,UT, a filament-prominence polar-crown configuration to the northwest accelerates its rise, transitioning into a CME  ($\Delta_{\rm A}\approx 6$\,h), with first signatures in SOHO/LASCO C2 after 20:48\,UT. There is some activity near AR\,11356, and the filament between ARs\,11353 and\,11354 continues to evolve, but neither is obviously connected to the erupting filament-prominence to the northwest.

%2011-Nov-30 : 5813
\begin{figure}
\epsscale{1.18}\plotone{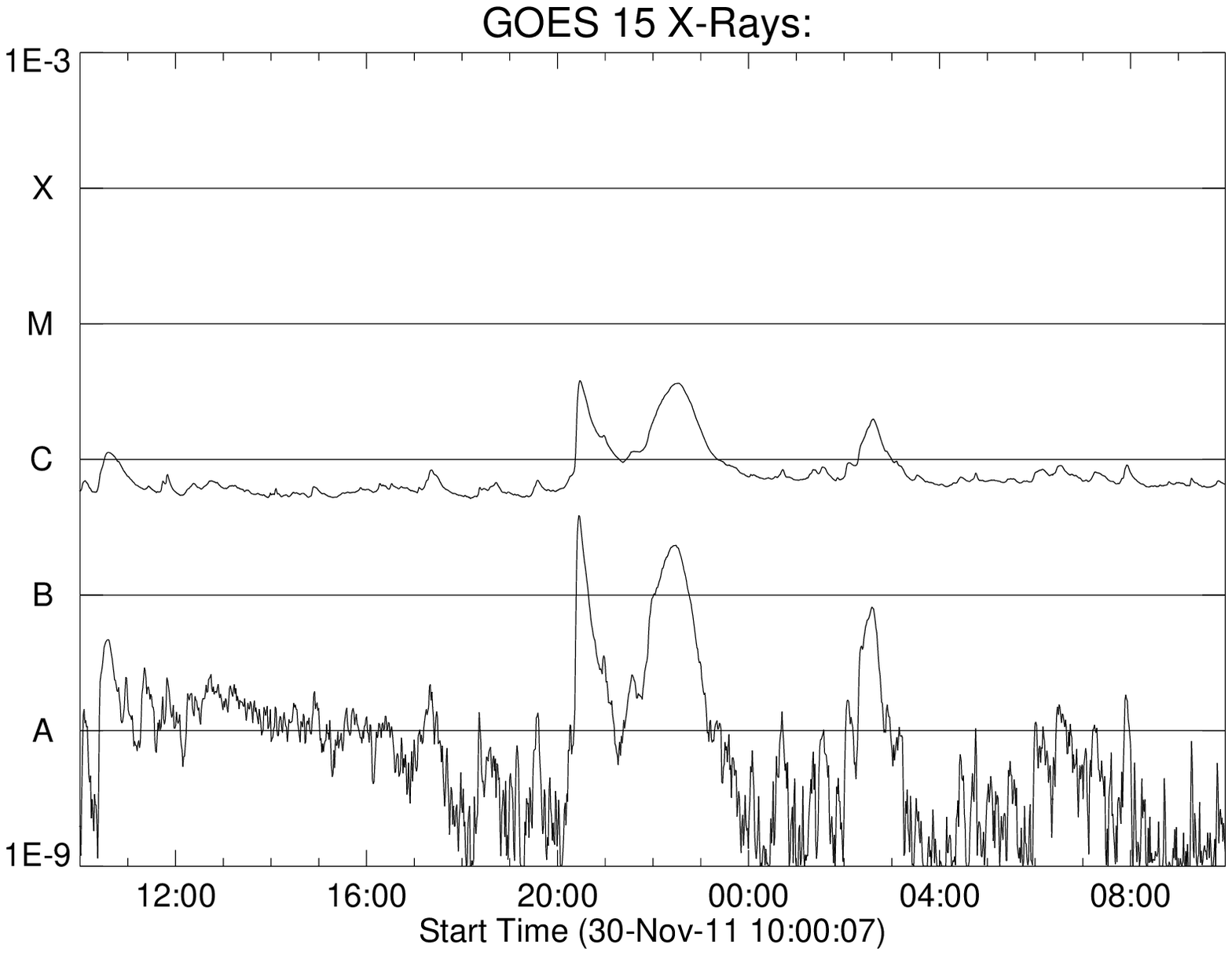}
\caption{GOES light curve for  case F, 2011/11/30.}\label{fig:goes20111130}
%\end{figure}
%\begin{figure}
\epsscale{1.18}\plotone{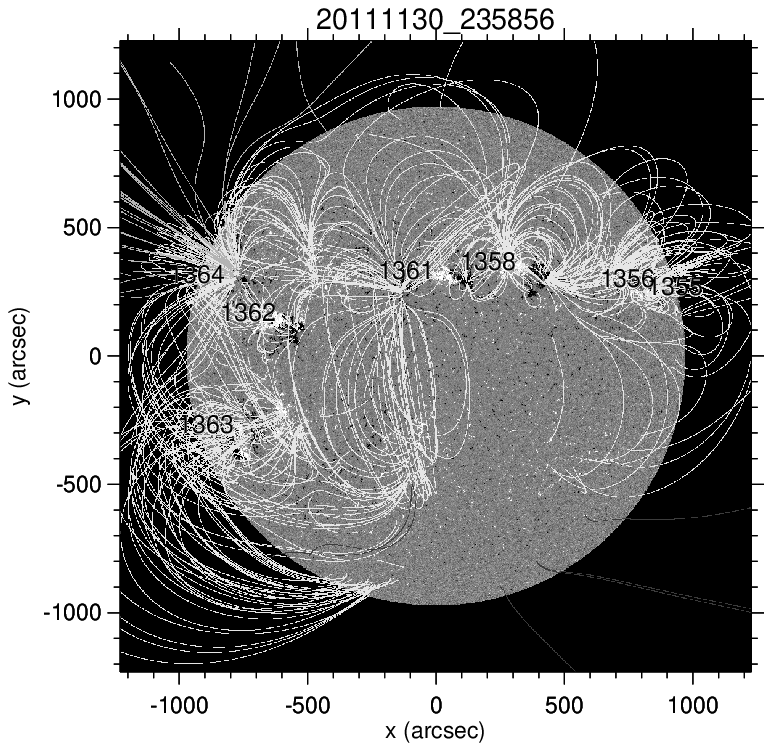}
\caption{HMI magnetogram, PFSS extrapolation, and AR numbers for case F,  2011/11/30.}\label{fig:pfss20111130}
\end{figure}
\subsection{Case F: 2011/11/30}
Around 2011/11/30 12\,UT a prominence over the east limb begins to rise. The prominence trails AR\,11362 which shows some brightening at that time, but without obvious GOES signatures above the background variability (Fig.~\ref{fig:goes20111130}). Around about 19\,UT, the prominence rise accelerates and transitions into an eruption  ($\Delta_{\rm A}\approx 7$\,h). On the other side of the Sun, a filament in the southern reaches of AR\,11355 begins to rise around 19:50\,UT, erupting after 20:25\,UT, in association with a mid-C class flare. The latter eruption occurs even as the off-limb corona to the east is still deforming in association with the prominence eruption. Around 21:40\,UT, the same filament configuration in AR\,11355 erupts once more  ($\Delta_{\rm A}\approx 5$\,h), into a C-class event of very comparable peak strength, but more gradual in both onset and decay.

Neither the  PFSS model (Fig.~\ref{fig:pfss20111130}) nor the MF field$^{\href{http://www.lmsal.com/~schryver/STYD/mfsnap5813.jpg}{S12}}$ show obvious connections between these two eruptions. The SOHO/LASCO  images show signatures of the two eruptions starting on the east limb around 19\,UT and from AR\,11355 in the west around 20:25\,UT entering the C2 field essentially simultaneously just after 21\,UT. The second eruption from AR\,11355 is a much brighter CME that becomes visible between 22:00\,UT and 22:36\,UT. In C3 images, the set of three eruptions looks like a single complex event, extending from east to west, with some brightness variations at all hour angles, giving this set of events the appearance of a single irregular halo CME. STEREO A/B COR2 data reveal that the southern component of this apparent halo event as seen from Earth perspective is, in fact, a slower moving CME that erupted from the far-side southern hemisphere (in an eruption starting around 18\,UT) even before the above eruptions started. We find no compelling evidence for a physical linkage between these three eruptions, although all three appear to occur from underneath the helmet streamer.

These events are an excellent demonstration that what appears as a single CME in coronagraph images, in fact is a composite event from two main eruptions, overlapping with even other events. There is no obvious evidence for causal couplings between these two eruptions within the low solar corona, however. 

\begin{figure}
\epsscale{1.18}\plotone{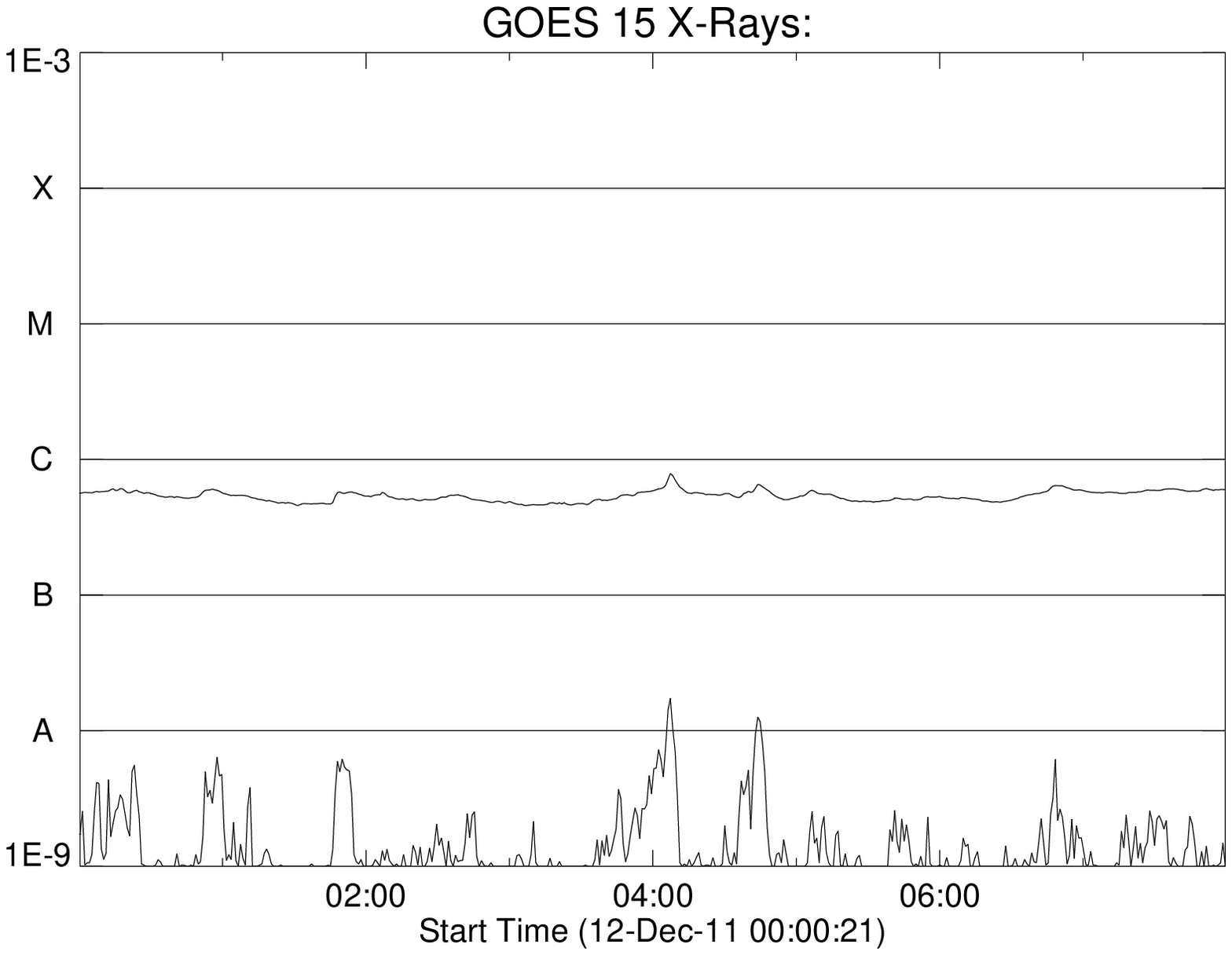}
\caption{GOES light curve for  case G, 2011/12/11.}\label{fig:goes20111211}
%\end{figure}
%\begin{figure}
\epsscale{1.18}\plotone{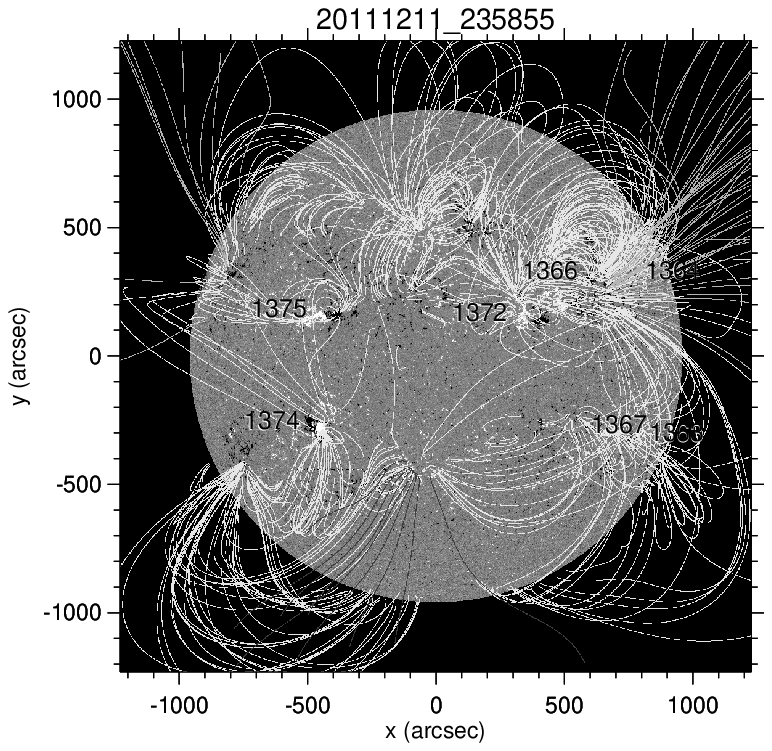}
\caption{HMI magnetogram, PFSS extrapolation, and AR numbers for case G,  2011/12/11.}\label{fig:pfss20111211}
\end{figure}
%2011-Dec-11 : 5824
%\clearpage
\subsection{Case G: 2011/12/11}
This case shows the eruption, starting around 5\,UT  ($\Delta_{\rm A}\approx 7$\,h), of a quiet-Sun filament (confirmed by H$\alpha$ observations by Meudon) with its enveloping rope configuration (clearly seen in the MF field model$^{\href{http://www.lmsal.com/~schryver/STYD/mfsnap5824.jpg}{S13}}$). During the course of the eruptions it is revealed that the quiet-Sun filament in fact connects to a trailing negative-polarity region some 60$^\circ$ away. Thus, it is a demonstration of a direct field connection some 60$^\circ$ trailing the initial eruption site that is evident only from field modeling, but that is not recognizable as such in a single coronal or H$\alpha$ image prior to the filament eruption. The destabilization, flare, and eruption of a region at the limb region (reaching no higher than B9) was a direct consequence of the eruption of a filament starting almost a solar radius away. Prior to that eruption, a high rope was seen to move out to about 1.2R$_\odot$ over the north-polar region, but without erupting into the SOHO/LASCO COR2 field of view; its relationship relative to the eruption and weak limb flare remains unknown. 

The long quiet-Sun filament configuration in the northeast quadrant destabilizes around 04\,UT. By 05:45\,UT, coronal brightenings are suggestive of eruption-driven reconnection (with a mild GOES increase at that time, but rising from about B7 to B8). In the subsequent 20 minutes, ribbon-like features and dimming regions develop, most prominently on the trailing side, that are commonly associated with eruptions of flux ropes. By 06:15\,UT, there are brightenings some 10 degrees eastward of the trailing end of the filament, revealing more extended field involvement. 

\begin{figure}
\epsscale{1.18}\plotone{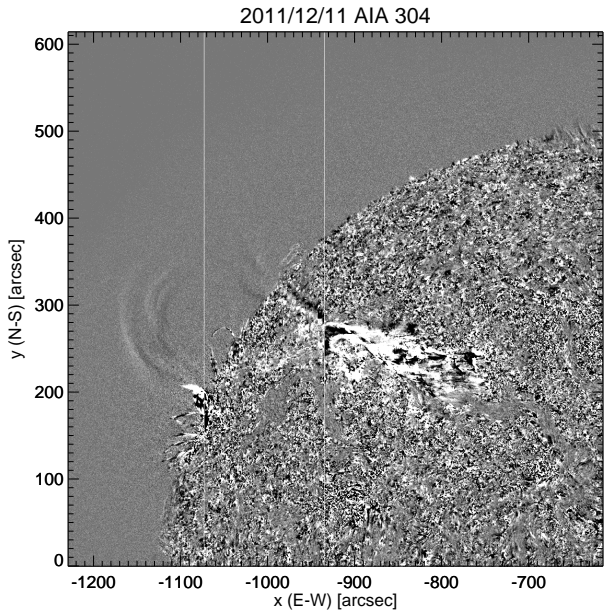}
\caption{Composite running-difference image for 304\,\AA\ exposures on 2011/12/11 (case G), displaying the northeast quadrant of the Sun (summarizing the image sequence shown in the on-line Table$^{\href{http://www.lmsal.com/forecast/STYD.html}{S0}}$ that also shows the STEREO images for comparison). This composite image highlights the connection between the on-disk and off-limb erupting regions. The figure is an assemblage of a series of running difference images, each computed for a time spacing 4\,min (the image sequence can be accessed on line$^{\href{http://www.lmsal.com/~schryver/STYD/rdiff20111211_304.mov}{S20}}$). To the right of the rightmost vertical white line is the difference image for 05:45\,UT, for the early phases of the eruption. To the left of the leftmost vertical line is the image for 07:07\,UT. Between the two vertical lines is a series of narrow strips cut out of running difference images with positions shifting from left to right with time, advancing in steps of 3\,min., moving in the general direction of motion of the eruption. }\label{fig:taffy304}
\end{figure}
The images from AIA's 304\,\AA\ channel most clearly reveal a direct coupling with a distant trailing region just at the east limb (Fig.~\ref{fig:taffy304}). The erupting filament is seen to have, or to develop, an extension that propagates towards the region at the limb. By 07\,UT, this filament extension is seen to distort the field in the region at the limb, pulling a segment of the field out of its initial configuration, stretching it high into the corona. Around 08\,UT, the distorted configuration of the limb region itself erupts (GOES 0.5-4.0\,\AA\ rising to about B9 starting at 08:05\,UT), with the last obvious signatures of that eruption ending around 08:45\,UT as the eastern end of the erupting filament exits the AIA field of view.

We note that around 08:20\,UT, a small filament leading the main erupting filament by about 40 degrees initiates an eruption  ($\Delta_{\rm A}\approx 5$\,h) that fully develops around 08:45\,UT. STEREO-A COR1 (then leading Earth by 107$^\circ$) shows a CME first developing at about 06:45\,UT, followed by a second at 09:10\,UT. SOHO/LASCO C2 images confirm that there are two distinct eruptions, so the timing of the eruption of the leading smaller filament around the end of the activity at the east limb appears not to be indicative of a single composite eruption. The bottom end of a U-shaped configuration detaches from the STEREO-A COR1 occulting disk at about 03:45\,UT, continuing a very gradual rise through at least 07:45\,UT when it fades from the running difference images at about 1.2R$_\odot$ above the solar surface. There is no obviously identifiable counterpart of this in SOHO/LASCO data, likely because the structure faded, if not stalled, before reaching the C2 field of view. It is possible that this reconfiguration is part of the overall set of events, either as another consequence or as an element of the subsequent eruption.

Prior to the primary eruption of the quiet-Sun filament, STEREO-A/B COR1 images show a gradual, faint eruption of a rope-like configuration already in progress at very high northerly latitudes shortly after midnight. 

The PFSS field shows two connected arcades: a long arcade over the erupting filament, and a side channel connecting to the region at the east limb. The MF field reveals a rope configuration that distributes its field between a bipolar area just south of the trailing end of the initially erupting filament (that in later magnetograms is shown to have a small active region injected into it while on the far side) and a negative-polarity region at the limb, that corresponds to the erupting limb region. The MF configuration for 2011/12/11 shows the low-lying flux rope (at and below 1.1 R$_\odot$) extending over 130 degrees, with a higher rope configuration (around $1.25-1.30$R$_\odot$) over a 70-degree span above the main filament eruption site; this high configuration may be what the STEREO A/B COR1 showed as a gradually rising rope, although then manifested at apparently higher latitudes. Six days later, the high rope has mostly disappeared from the model, having evolved away while the low-lying rope has lost its leading segment around where the eruption occurred, both presumably in part because of newly emerged field on the far side, and possibly because of a loss of equilibrium in the overall configuration. 

In the eruption evolving around 8\,UT, the 304\,\AA\ image sequences (see the on-line SDO/STEREO 304\,\AA\ movie$^{\href{http://www.lmsal.com/forecast/STYD.html}{S0}}$ and, in particular, the running difference image set$^{\href{http://www.lmsal.com/~schryver/STYD/rdiff20111211_304.mov}{S20}}$  described in the caption to Fig.~\ref{fig:taffy304}) are highly suggestive of a field connection that the field models suggested possible, but did not themselves directly contain. With these events occurring at the east limb, however, no model could have captured this well given the unavailability of up-to-date magnetic information for the far hemisphere of the Sun as seen from Earth.

\begin{figure}
\epsscale{1.18}\plotone{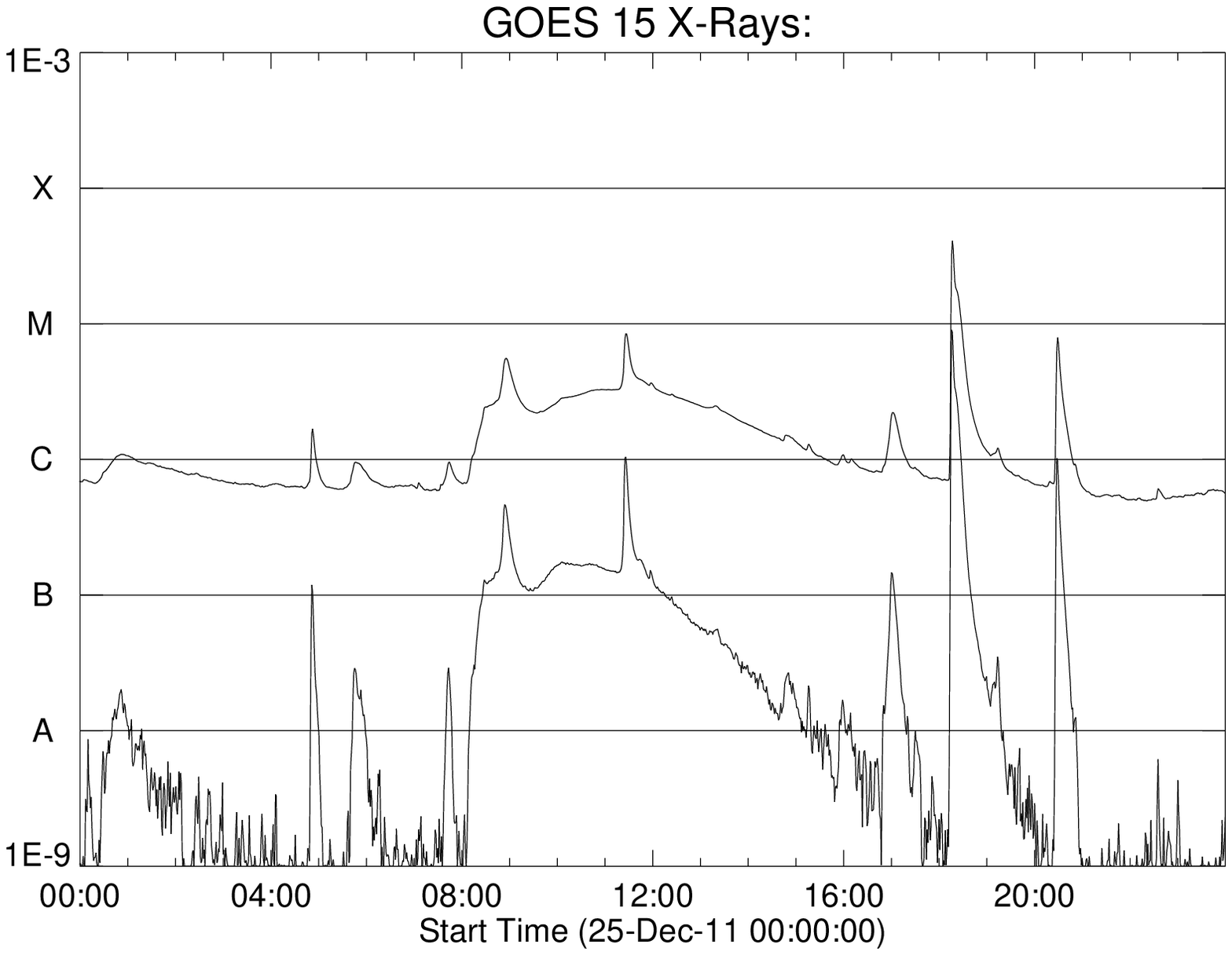}
%\end{figure}
%\begin{figure}
\epsscale{1.18}\plotone{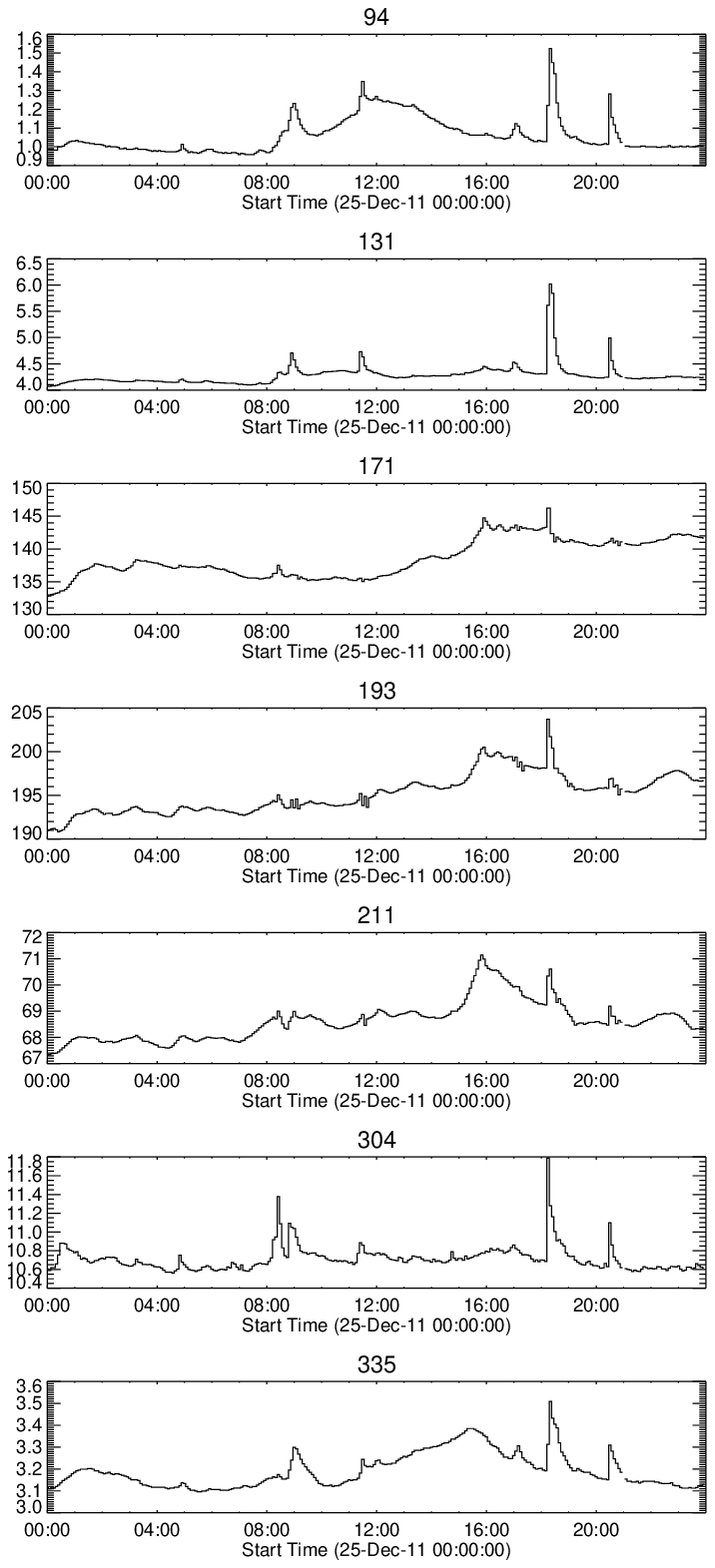}
\caption{GOES and AIA light curves for case H,  2011/12/25  (for AIA in DN/pixel/s).}\label{fig:goes20111225}\label{fig:lcurve20111225}
\end{figure}
\begin{figure}
\epsscale{1.18}\plotone{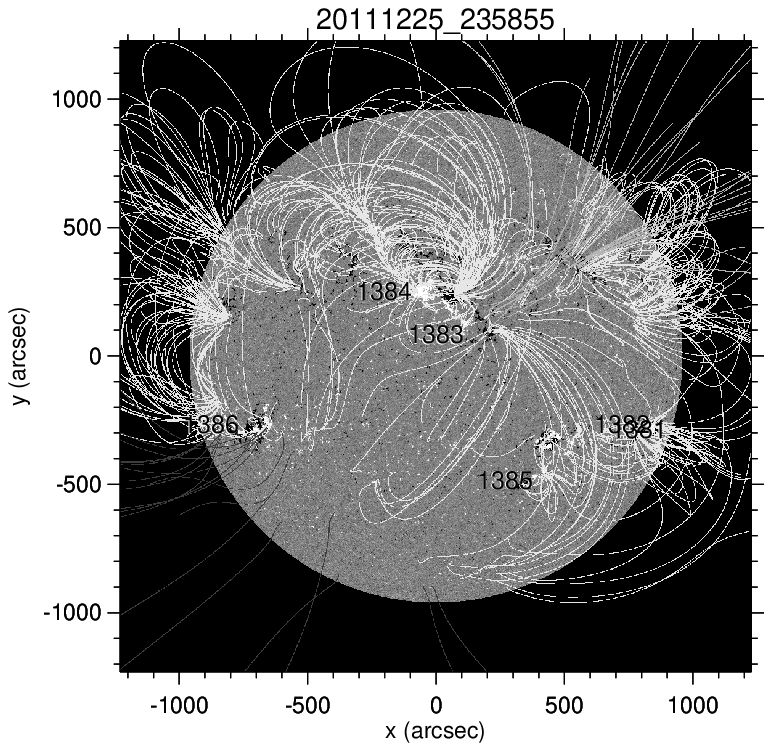}
\caption{HMI magnetogram, PFSS extrapolation, and AR numbers for case H,  2011/12/25.}\label{fig:pfss20111225}
\end{figure}
\begin{figure}
\epsscale{1.18}\plotone{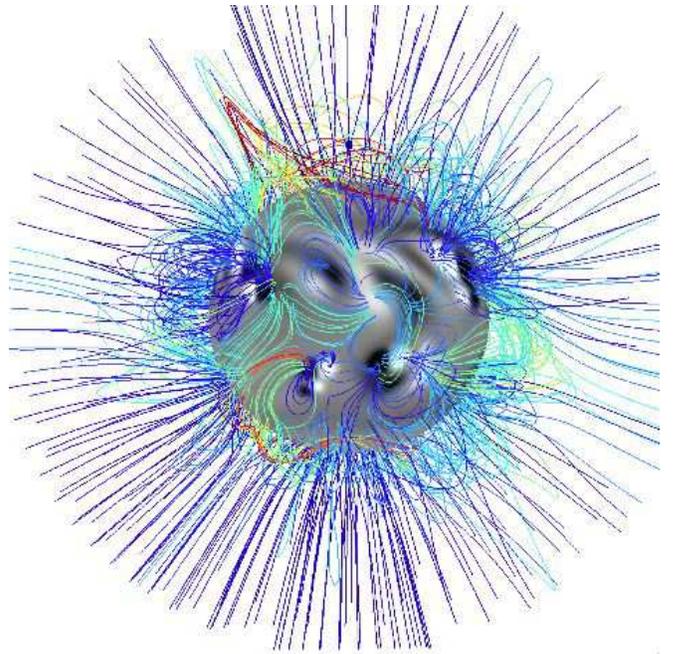}
\caption{Magnetofrictional field model for  case H, 2011/12/25, rotated eastward by 30 degrees relative to the Carrington longitude  (cf. Fig.~\ref{fig:mf20110216} for a description of the details).}\label{fig:mf20111225}
\end{figure}
\begin{figure}
%\epsscale{1.18}\plotone{mf20111225_ropes.eps}
\epsscale{1.18}\plotone{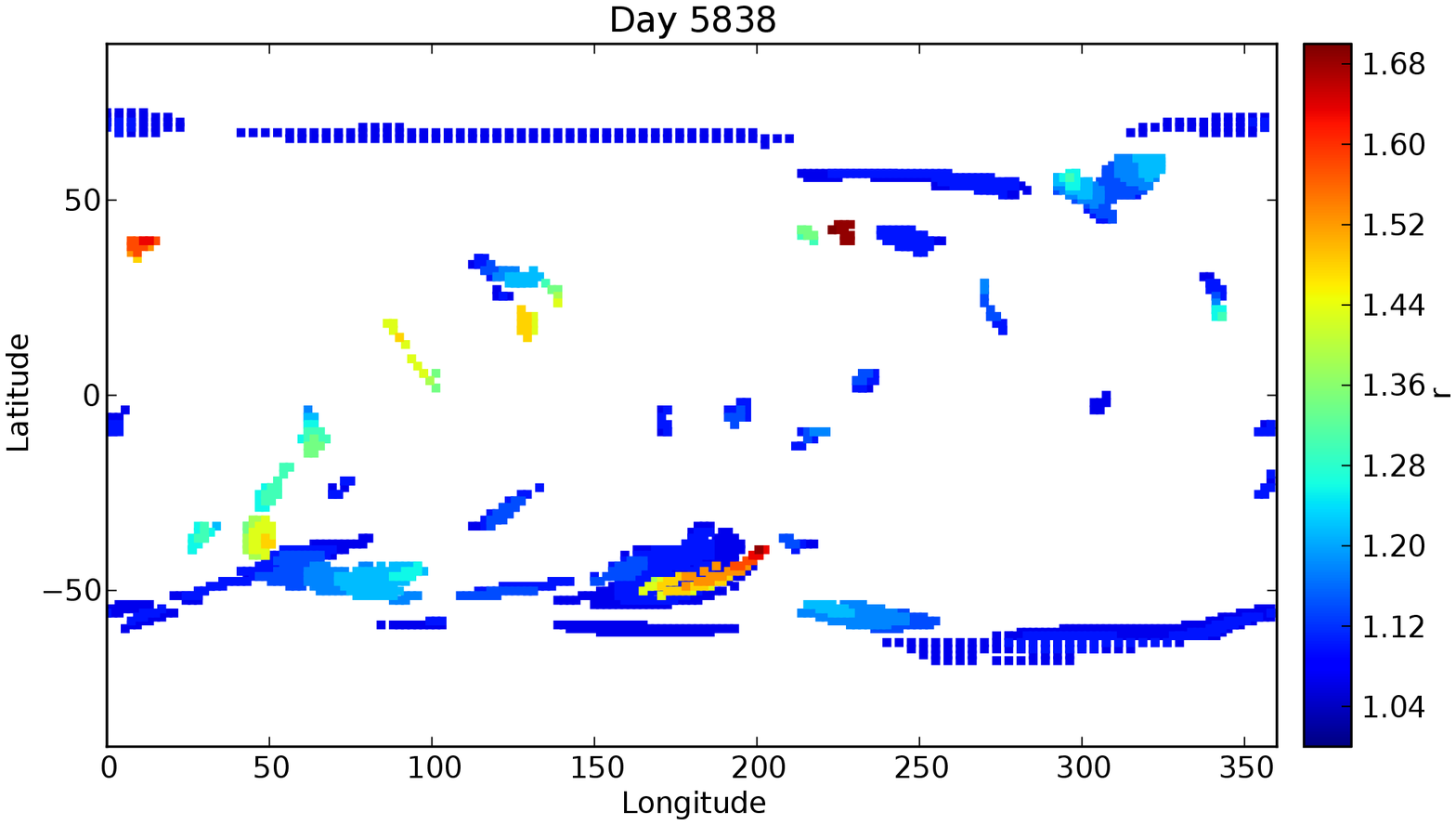}
\caption{Flux rope positions, colored by height (see color bar, in units of solar radii), from the magnetofrictional model for  case H. The Carrington longitude of disk center on 2011/12/25 is 209$^\circ$}\label{fig:mf20111225_ropes}
\end{figure}
%2011-Dec-25 : 5838
\subsection{Case H: 2011/12/25}
Early on 2011/12/25, two quiet-Sun filaments erupt from near central meridian (leading ARs\,11384 and\,11383 by about 30 degrees, cf., Fig.~\ref{fig:pfss20111225}) in the northern hemisphere.  The first, northernmost filament is in early eruption at 2011/12/24 23:45UT, with ribbons forming around 2011/12/25 00:15\,UT  ($\Delta_{\rm A}\ga 8$\,h). The erupting filament rapidly rises, exiting the AIA field of view by 01\,UT, leaving behind a post-eruption arcade that relaxes at least until after 06\,UT. The southernmost signal that we see in AIA running-ratio movies$^{\href{http://www.lmsal.com/~schryver/STYD/AIAtriratio-211-193-171-2011-11-25T0000.mov}{S7}.\href{http://www.lmsal.com/~schryver/STYD/AIAtriratio-211-193-171-2011-11-25T0600.mov}{S8}}$ extend southward of the next event: the second filament, just to the south of it, responds gradually starting from about 03\,UT, accelerating from about 05:30\,UT onward, with reconnection signatures starting around 07\,UT.

Around 07:50\,UT, another eruption begins, this time from a region just over the northwest limb. It forces its way through the coronal field starting from 08:25\,UT onward  ($\Delta_{\rm A}\approx 3$\,h). Starting at 08:45\,UT AR\,11385 flares ($\approx$C7) and erupts, associated with a coronal propagating front, and once more ($\approx$C9) around 11:15\,UT. In the GOES light curves, all of the above events form a single curve with two C-spikes on it (Fig.~\ref{fig:goes20111225}). 

At 18:11\,UT, an M4.0 starts from AR\,11385, associated with a very pronounced, far-reaching coronal propagating front  ($\Delta_{\rm A}\approx 1$\,h inside the region, and approximately 3\, in its surroundings). As that front reaches the trailing AR\,11386, just after 18:30\,UT, a relatively  minor eruption occurs within it. Connections between these regions were suggested earlier in the day when, in association with the 08:45\,UT flare in AR\,11385, brightenings in AR\,11386 were observed, and again in events of different magnitude at 17:05\,UT and 18:15\,UT, interspersed with others in which no such synchronicity was observed.

The PFSS (Fig.~\ref{fig:pfss20111225}) and MF$^{\href{http://www.lmsal.com/~schryver/STYD/mfsnap5838.jpg}{S14}}$ models do not reveal field connections or topological relationships that might connect the various eruptions of the on-disk filaments with the over-the-limb configuration, with AR\,11385, or of AR\,11385 with AR\,11386.  
There are several high ropes (fragments) in the magnetofrictional model for this date (Fig.~\ref{fig:mf20111225_ropes}): one between 1.25 and 1.6$R_\odot$ between ARs\,11385 and 11386, overlying a much lower one below 1.1$R_\odot$, and above 1.6$R_\odot$ overlying the two erupting filaments.

We note that whereas the GOES light curve in Fig.~\ref{fig:goes20111225} suggests that the eruptions form a long-duration event in which energy conversion occurs over some 24\,h before the corona relaxes to its pre-events emission levels. AIA's light curves in that figure show a similar behavior for the shortest wavelength channel at 94\,\AA, which has a strong contribution from Fe\,XVIII lines. The channels responsive to lower coronal temperatures, specifically 171, 193, and 211\,\AA, show a strong peak developing from about 15\,UT onward, which has only a very weak counterpart in the GOES light curves. These signals are dominated by post-eruption emission from an event that started behind the northwestern limb, of which the post-eruption loops are showing up over the limb in the AIA observations  ($\Delta_{\rm A}\ga 10$\,h).

We propose that the eruptions of the on-disk filaments may be coupled as in the mechanism explored by \citet{2011ApJ...739L..63T}, and that the propagating front emanating from the M4.0 flare in AR\,11385 was instrumental in upsetting the balance of the field in AR\,11386 or at least in accelerating the destabilization. We find no evidence for direct field couplings in the equilibrium field models, however, between the on-disk filament eruptions and the over-the-limb eruption, or between those and the subsequent mid-C class flaring in AR\,11386.

\begin{figure}
\epsscale{1.18}\plotone{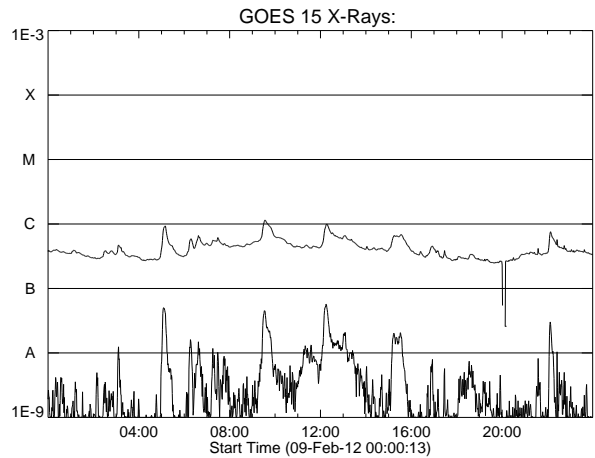}
\caption{GOES light curve for  case I, 2012/02/09.}\label{fig:goes20120209}
\end{figure}
\begin{figure}
\epsscale{1.18}\plotone{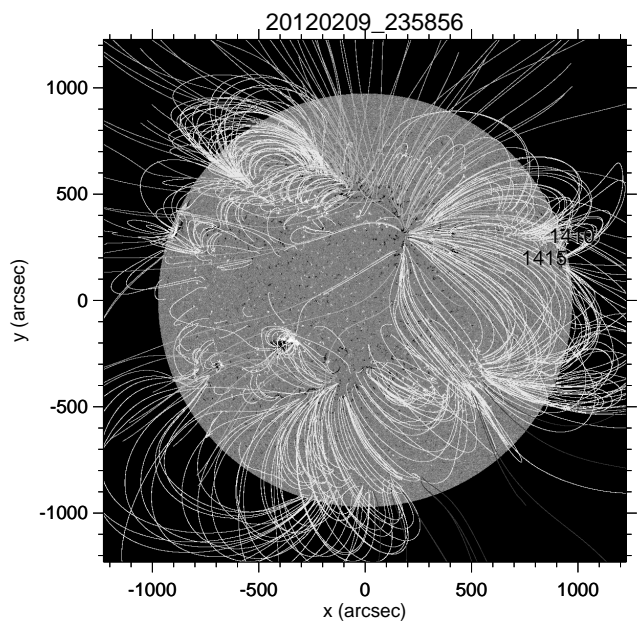}
\caption{HMI magnetogram, PFSS extrapolation, and AR numbers for case I,  2012/02/09.}\label{fig:pfss20120209}
\end{figure}
%2012-Feb-9    : 5884
\subsection{Case I: 2012/02/09}
Around 2012/02/09 07\,UT changes begin to occur in a quiet-Sun region around $(x,y)=(-300,+400)$\,arcsec from disk center. These changes suggest an eruption, with intensity signatures coupling in a compact, unnumbered, emerging bipolar region in the southern hemisphere around $(x,y)=(-300,-200)$. No obviously related additional activity occurs in that emerging region as a result of this, although some activity is ongoing in association with its emergence throughout the day. The thermal evolution of the erupted region in the north continues until at least 12\,UT  ($\Delta_{\rm A}\approx 5$\,h). 

Starting around 17\,UT, brightness and thermal changes are seen in that erupted region around $(x,y)=(-300,+400)$  ($\Delta_{\rm A}\approx 7$\,h), which continues as a filament/prominence due east at the limb begins to erupt, accelerating into a CME around 19\,UT (visible in SOHO/LASCO C2 after 20:48\,UT), together with a quiet-Sun filament reconfiguration to the southeast of, and connected with, the bipolar region at $(x,y)=(-300,-200)$  ($\Delta_{\rm A}\approx 8$\,h). As this joint eruption continues, the regions in the vicinity of $(x,y)=(-200,-200)$ to $(x,y)=(-200,+400)$ exhibit a pronounced change, starting with an apparent cooling (dominant 171\,\AA\ channel) from 20\,UT onward, then recovering by the end of the day.  The GOES light curve reaches C level only twice, and then briefly, during these events (Fig.~\ref{fig:goes20120209}).

The PFSS model (Fig.~\ref{fig:pfss20120209}) does not show obvious connections between these regions of activity, and neither does the MF$^{\href{http://www.lmsal.com/~schryver/STYD/mfsnap5884.jpg}{S15}}$ model field. SOHO/LASCO C2 data suggest,  however, that the prominence eruption from the northeast limb region is coupled to regions to a clock angle of about 08h, i.e., to the general direction of the emerging bipolar region in the southern hemisphere and the nearby quiet-Sun filament, even deforming the streamer slightly to the south of that by its expansion. The absence of clear stretched-field signatures, other than the expanding CME (shock) front, in STEREO/A COR1, however, suggests that the structures seen from SOHO/LASCO C2 may not, in fact, map onto the disk as seen from Earth.

There is no obvious direct evidence for coupled substantial activity for 2012/02/09, although the pronounced cooling seen in the quiet-Sun filament region around $(x,y)=(-300,+400)$ during the rise and eruption of the prominence over the northeast limb is suggestive of the existence of long-range couplings in the field that are manifested during the eruption possibly through high-arching field next to open field that is rooted in a patch around $(+100,+200)$. We note that with these events occurring on the eastern hemisphere, and major parts of that on the east limb, the model field configuration may not reflect all the relevant parts as the field underlying these events is far from fully assimilated into the field models. Another interpretation is the analogy with the  model discussed by \citet{2012ApJ...759...68L}, in which reconnection during the eruption process forms long-range connections that can lead to dimming and adiabatic cooling of plasma on loops connecting far away.

\section{In summary}\label{sec:conclusions}
We have shown evidence that directly-connected, adjacent coronal volumes (in any combination of active and quiet regions) can be involved in sympathetic events when the connecting field is affected by a flare or eruption (examples discussed in this study include cases A, B, C, G). Similarly, adjacent filaments, nested within a larger overall configuration, can be associated in coupled eruptions, as explored in the MHD model by \citet{2011ApJ...739L..63T}, for which cases A and H contain good illustrations. 

For case~I, there is only weak evidence of long-range couplings in the PFSS and MF field models between an erupting east-limb filament and a dimming, cooling region in the quiet Sun on disk. This coupling appears to be through high-arching field under a helmet adjacent to an open-field patch (reminiscent of the model by Lugaz et al, 2012). Although the field configuration for such a volume of the corona near the eastern limb is subject to major uncertainties, the unusual evolution of intensity profiles and thermal signatures over an  on-disk region of quiet-Sun as an east-limb filament eruption proceeds suggests that such connections do, in fact, exist. We suggest that these eastern-hemispheric events illustrate that magnetic connections in this case are not correctly approximated by the PFSS and MF field models simply because much of the surface field involved is not, or only poorly, visible by existing magnetographs.

Another illustration of the problems associated with the poor magnetograph coverage is found in case~B: a magnetofrictional field model for 2011/02/16 reveals a set of connecting field structures that ties together all of the major activity for the preceding days of activity, but these connections had not fully developed in the model by then. Given the patterns of coronal activity on the 14th and 15th, it is  possible that connections that already exist in the real corona were yet to develop in the model. 

Another likely example of the consequences of not having all of the information of the surface field near the edge of the visible disk and beyond is that of case~G, where the AIA observations show a direct connection between two distant regions linked by a coupled eruption and flare. A fourth example of this is that of case~A described in detail by \citet{schrijver+title2010} who explicitly show the substantial coronal evolution between a PFSS model that includes regions that emerged on the far solar hemisphere and one that does not.

In other cases, field models are helpful in revealing connections where nestings of field, some including ropes, could not evolve without upsetting adjacent fields (of which case~C is a good example). In other cases, these field models reveal that evolution in one region affects the evolution in a distant region via connections that sit side by side in a third location (e.g., cases~B and~D) and that thus may involve a coronal null. Detailed MHD modeling is needed to reveal to what degree such multi-hop connections, mutually influencing each other as they evolve,  communicate evolutionary signals leading to sympathetic activity. 

Among the cases discussed, we have identified a few in which coronal propagating expansion fronts or shock waves (or both) reach (the vicinity of) other regions (quiet or active) that then destabilize and erupt (such as in cases~C and~H, and perhaps B). Understanding the potential couplings in such cases also calls for detailed evolutionary MHD models.

In other event sets, such as some in case~H and for case~F, there may be no evidence for connections of any kind between two distant eruptions. In those cases distinct eruptions from different sites overlap in time so that the resulting inner-heliospheric perturbation develops as a single composite event. Knowing of such composite events is important to understanding the possible impacts on geospace and to the forecasting of such impacts. In contrast to this, case~E shows how much of the front-side corona can be evolving, but without introducing noticeable confusion about the source regions and triggering events of inner-heliospheric perturbations. 

\section{Discussion and conclusions}\label{sec:discussion}
Our study leads us to two primary conclusions: (1) long-range couplings by several distinct physical mechanisms affect the evolution of the large-scale corona, and (2) the structure of the large-scale coronal field is the product of insertion, buildup, ejection, and relaxation of electrical current systems that are inadequately represented by present-day modeling and impossible to reconstruct from present-day observational data. These properties of the dynamic, 3-dimensional, dynamic corona need to be incorporated into assimilative models that couple the solar surface to the heliosphere to accurately forecast the solar wind,  its variable properties, and the pathways of solar energetic particles. 

Present-day instrumentation supported by relatively elementary modeling for the global coronal field has revealed at least three different ways eruptive and explosive events in the solar outer atmosphere influence the destabilization of other regions: evolving direct magnetic coupling, distortion of the enveloping field by a large eruption developing into a CME, and the effects of an expansion front or coronal wave associated with a major flare or eruption. Observations suggest that indirect coupling through an intermediate region adjacent to a mutual separatrix surface could be another pathway for couplings to occur.

Whereas in some cases the evidence for coupling between events is direct and uncontroversial, in many others it depends on the applicability of the model fields used in the interpretation. This presents a major problem for several reasons. First, the overall coronal field is built up through flux emergence, shear, and eruptions that require a full-sphere model and likely at least a full rotation of model time prior to the event to be studied \citep[see the model and discussion by][]{yeates+etal2008}. It may require more than a year of solar time for the high-latitude fields that involve, among others, the polar crown filament configurations which are long-term assemblages of years of active-region decay products \citep[e.g.,][]{2012ApJ...753L..34Y}. This is not only computationally demanding, but also requires that we increase our observational coverage of the solar surface in at least line-of-sight, perhaps even vector-magnetograms, from only the Earth-facing side as available at present to a substantially larger coverage of the solar surface. 

In order to succeed in obtaining an acceptable model representation of the real-world coronal field,  we have to realize that field disturbed by a CME requires a good fraction of a day to relax. A lower limit to that time scale is set by 
the EUV afterglow of  post-eruption arcades \citep[extending beyond, for example, the time interval during which supra-arcade downflows might be visible, e.g.][]{2012ApJ...747L..40S}.
For 21 eruptions from outside active-region core domains studied here, the average coronal relaxation time estimated from 171\,\AA, 193\,\AA, and 211\,\AA\ signals is  $\langle \Delta_{\rm A,QS} \rangle = (6.9 \pm 3.7)$\,h.
 
The time scale for the EUV afterglow following quiet-Sun eruptions of  $\langle \Delta_{\rm A,QS} \rangle = (6.9 \pm 3.7)$\,h is much longer than the corresponding signatures of flares and eruptions within active regions. There $\langle \Delta_{\rm A,AR} \rangle = (0.6 \pm 0.5)$\,h (estimated from 8 events, with one strong outlier). We can compare that ratio of  $\langle \Delta_{\rm A,QS} \rangle / \langle \Delta_{\rm A,AR} \rangle\approx 12$ with a ratio of typical timescales involved. For length scales $\ell_{\rm QS,AR}$ and Alv{\'e}n speeds $v_{\rm QS,AR}$ we can derive a time scale ratio of $\tau_{\rm QS}/\tau_{\rm AR} = (\ell_{\rm QS}/v_{\rm QS}) / (\ell_{\rm AR}/v_{\rm AR}) = 
(\ell_{\rm QS}/\ell_{\rm AR}) (\rho_{\rm QS}/\rho_{\rm AR})^{1/2}/((B_{\rm QS}/B_{\rm AR}) $). For typical length scales of 250\,Mm and 50\,Mm, field strengths of 10\,G and 100\,G, and densities of $10^8$\,cm$^{-3}$ and $10^9$\,cm$^{-3}$ for quiet-Sun coronal regions and active-region flaring interiors, we find $\tau_{\rm QS}/\tau_{\rm AR} \approx 16$. That ratio is compatible with the observed ratio of time scales, consistent with the interpretation that the EUV eruption afterglow in quiet Sun is a signature of the post-eruption reconnection like the relaxation that occurs in active regions after eruptions. 

With 4 CMEs going off on an average day, this long reconnection time scale means that the high coronal field is most likely often evolving from the disturbing effects of previous eruptions at the time of any event that we elect to study.
How much of the coronal field is involved in such reconfiguration? One might look at the angular extent of CMEs to estimate this. \citet{robbrecht+etal2009} show a power-law distribution of CME opening angles with an average power-law index of $-1.66$ that holds for opening angles of 10$^\circ$ up to above 120$^\circ$.  The average opening angle is about 50$^\circ$, corresponding to about 1/10th of the full sky as seen from the Sun. One might thus infer that roughly 1/10th of the corona by volume is involved in an average eruption. However, even a CME with a small opening angle must break all field that closes over its site of origin. With a typical base cross section of order 100$^\circ$ for the helmet streamer belt in the PFSS approximation, and assuming a comparable width along and across the helmet direction, then up to $\sim 1/3$rd or the solar surface may have a fraction of its field forced to reconnect during a CME. The latter estimate is consistent with the extent over which perturbations are seen to travel in the tri-ratio movies such as shown in the on-line Table$^{\href{http://www.lmsal.com/forecast/STYD.html}{S0}}$. If we take $\langle \Delta_{\rm A,QS}\rangle$ to be characteristic of the reconnection time scale for the high coronal field, and for an average of 4 CMEs/day that each reach over about 1/3rd of the solar surface, about one in $1/(1-(11/12)^3)\approx 4$ of all CMEs affect field that is still relaxing from preceding CMEs. The time scale for ongoing reconnection is likely larger than $\langle \Delta_{\rm A,QS}\rangle$; if we use a reconnection time scale of $2\langle \Delta_{\rm A,QS}\rangle$ then close to one in $1/(1-(5/6)^3)\approx 2$ CMEs occur while the field that it encompasses is still relaxing. 

In view of this, one should question the validity of any global coronal model at the start of a selected event that does not include in its computation at least several preceding eruptions. The events of, and model for, case~A are an excellent illustration of this: even as the model shows how one filament can be destabilized in the wake of another's eruption, and although this certainly appears to be the case on that date, the second filament does not take off until half a day after the nearby afterglow of the reconnection arcades associated with the first eruption faded away; only the continuing afterglow in a more distant location (at the edge of the northern coronal hole) bore witness to the ongoing deformation of the field enveloping the filaments and flux-rope configurations. The double filament eruptions in case~H are similar in that respect, with the eruptions in that case separated by approximately 7\,h.

This work has illustrated that we should anticipate that in many instances, connections in the solar corona exist that present-day coronal field models do not reveal. The fundamental problem that we face in understanding the connections in the solar corona even in the absence of eruptive events is that of the dependence of its instantaneous state on  what happened before, i.e., its magnetohydrodynamic hysteresis. In addition to the effects on time scale from flux emergence in days to field evolution over months, there are those that occur on time scales of hours to a day associated with CMEs. These are particularly important to understanding the evolution, and at least in some cases the triggering, of CMEs as well as the generation and escape pathways for solar energetic particles into the heliosphere \citep[see also, for example, the study by][and references therein on the role of precursor activity in seeding the coronal environment with particles that can be effectively accelerated to high energies in subsequent events]{2013ApJ...763..114S}. 

As we are faced with inadequate observational coverage and consequent model abilities, it remains difficult to establish how often the physical coupling of one event with another impacts coronal evolution. The importance for flare/eruption forecasting and thereby space-weather forecasting therefore remains unquantified. The evidence presented in this study motivates development of advanced capabilities and further study of the available data sets. 
Over the years, MHD and PFSS field models \citep[see, e.g.,][for a comparison between these]{riley+etal2006ApJ} have demonstrated that the largest-scale coronal configuration is represented fairly well: coronal hole boundaries and even their dynamics are approximated fairly well by open field regions in global PFSS models \citep{wang+sheeley93}, solar wind models based on series of PFSS models are reasonable approximations of the quiescent solar wind particularly in view of our limited observational coverage of the solar surface magnetic field \citep{arge+pizzo2000}, coronal streamers modeled for eclipses approximate some of the observed ones well \citep{2010A&A...513A..45R}, and even a fly-through of comet Lovejoy shows reasonable agreement with an MHD model \citep{lovejoymhd2013}. On the other hand, each of these tests of our understanding of the large-scale coronal structure and its evolution exhibits substantial mismatches \citep[see, e.g.,][and references therein for discussions of these aspects]{2011SoPh..274..361R,2012LRSP....9....6M}: modeled open-field regions do differ substantially from observed coronal-hole outlines, solar wind predictions are far from perfect in both speed and field polarity, the model streamer structures do not always reflect the directions of observed streamers differ while some observed helmet structures are missed altogether, the tail dynamics of comet Lovejoy suggests mismatches in field patterns of at least several degrees, and even the evolution of the Sun's open flux over time remains unexplained to within a range of a factor of at least two. All of these mismatches are likely caused by a mixture of the incomplete knowledge of the surface magnetic field, the evolving patterns of the coronal electromagnetic field that continually evolves following flux emergence and decay as well as eruptive events, and of assumptions and approximations made in the modeling. Numerical experiments can help us understand the relative weighting of the impacts of observational and modeling limitations, but ultimately sustained observations from perspectives well away from the Sun-Earth line, of the high corona, and of the details of current buildup and disappearance are essential in making the next leap forward in understanding the corona and the space weather that it drives. 

\acknowledgements
We thank Nariaki Nitta, Bernhard Fleck, and Miho Janvier for discussions of early versions of the manuscript.  
This work was supported by NASA contract NNG04EA00C
in support of the SDO Atmospheric Imaging Assembly.

\references

\end{document}